\documentclass[prb,twocolumn,nopacs,floatfix,superscriptaddress]{revtex4} 

\usepackage{amssymb}
\usepackage{amsmath}       
\usepackage{graphicx}
\usepackage{hyperref}
\usepackage{color}
\usepackage{enumerate}
\usepackage[normalem]{ulem}

\begin{document}

\clubpenalty=10000
\widowpenalty=10000
\brokenpenalty=10000
\tolerance=9000
\hyphenpenalty=10000

\title{\boldmath Magnetization-polarization cross-control near room temperature in hexaferrite single crystals \unboldmath}

\author{V. Kocsis}
\affiliation{RIKEN Center for Emergent Matter Science (CEMS), Wako, Saitama 351-0198, Japan}

\author{T. Nakajima}
\affiliation{RIKEN Center for Emergent Matter Science (CEMS), Wako, Saitama 351-0198, Japan}

\author{M. Matsuda}
\affiliation{Neutron Scattering Division, Oak Ridge National Laboratory, Oak Ridge, Tennessee 37831, USA}

\author{A. Kikkawa}
\affiliation{RIKEN Center for Emergent Matter Science (CEMS), Wako, Saitama 351-0198, Japan}

\author{Y. Kaneko}
\affiliation{RIKEN Center for Emergent Matter Science (CEMS), Wako, Saitama 351-0198, Japan}

\author{J. Takashima}
\affiliation{RIKEN Center for Emergent Matter Science (CEMS), Wako, Saitama 351-0198, Japan}
\affiliation{Engineering R $\&$ D Group, NGK SPARK PLUG CO., LTD. Minato-ku, Tokyo 108-8601, Japan }

\author{K. Kakurai}
\affiliation{RIKEN Center for Emergent Matter Science (CEMS), Wako, Saitama 351-0198, Japan}
\affiliation{Neutron Science and Technology Center, Comprehensive Research Organization for Science and Society (CROSS), Tokai, Ibaraki 319-1106, Japan}

\author{T. Arima}
\affiliation{RIKEN Center for Emergent Matter Science (CEMS), Wako, Saitama 351-0198, Japan}
\affiliation{Department of Advanced Materials Science, University of Tokyo, Kashiwa 277-8561, Japan}

\author{F. Kagawa}
\affiliation{RIKEN Center for Emergent Matter Science (CEMS), Wako, Saitama 351-0198, Japan}
\affiliation{Department of Applied Physics, University of Tokyo, Hongo, Tokyo 113-8656, Japan}

\author{Y. Tokunaga}
\affiliation{RIKEN Center for Emergent Matter Science (CEMS), Wako, Saitama 351-0198, Japan}
\affiliation{Department of Advanced Materials Science, University of Tokyo, Kashiwa 277-8561, Japan}

\author{Y. Tokura}
\affiliation{RIKEN Center for Emergent Matter Science (CEMS), Wako, Saitama 351-0198, Japan}
\affiliation{Department of Applied Physics, University of Tokyo, Hongo, Tokyo 113-8656, Japan}

\author{Y. Taguchi}
\affiliation{RIKEN Center for Emergent Matter Science (CEMS), Wako, Saitama 351-0198, Japan}

\begin{abstract}
Mutual control of the electricity and magnetism in terms of magnetic ($H$) and electric ($E$) fields, the magnetoelectric (ME) effect, offers versatile low power-comsumption alternatives to current data storage, logic gate, and spintronic devices~\cite{Kimura2003,Hur2004,Matsukura2015NatNano,Fusil2014}.
Despite its importance, $E$-field control over magnetization ($M$) with significant magnitude was observed only at low temperatures~\cite{Tokunaga2012NatPhys,Chai2014,Zhai2017NatMat,Chu2008,Heron2014a}.
Here we have successfully stabilized a simultaneously ferrimagnetic and ferroelectric phase in a Y-type hexaferrite single crystal up to 450\,K, and demonstrated the reversal of large non-volatile $M$ by $E$ field close to room temperature.
Manipulation of the magnetic domains by $E$ field is directly visualized at room temperature by using magnetic force microscopy.
The present achievement provides an important step towards the application of ME multiferroics.
\end{abstract}


\maketitle

Multiferroic materials, endowed with both orders of polarization ($P$) and $M$, exhibit various intriguing phenomena due to the interplay of magnetic and electric degrees of freedom, such as $H$-induced $P$ flop~\cite{Kimura2003,Hur2004}, $E$-control of magnetic helicity~\cite{Yamasaki2007,Tokunaga2010PRL}, and optical non-reciprocal directional dichroism~\cite{Arima2008JPCM,Kezsmarki2011}.
The cross-coupling phenomena can greatly expand the functions of materials, and hence the multiferroic materials are anticipated to be applied to technological devices.
In particular, non-volatile, $E$-driven reversal of $M$ without significant dissipation will lead to magnetic memory devices with ultra-low power-consumption. 

To accomplish this goal, strong coupling between $P$ and $M$ is necessary.
In general, depending on the microscopic mechanism of $P$ generation, the strength of the cross-coupling is different:
In type-I multiferroics where $P$ emerges potentially at a high temperature, but independently of the magnetic ordering, the coupling between $P$ and $M$ is weak, while in type-II multiferroics where $P$ is induced by magnetic ordering, the coupling between $P$ and $M$ is strong~\cite{Khomskii2009}.
Thus far, the $E$-induced $M$ reversal has been investigated for both type-I and type-II mutiferroics.
Heterostructures based on BiFeO$_3$ belonging to the type-I category have been demonstrated to be promising~\cite{Chu2008,Heron2014a,Sosnowska2013,Hojo2017}, while $H$-induced $P$ reversal is difficult~\cite{Tokunaga2015}.
In the type-II category, good performance has been reported for hexaferrite materials~\cite{Kimura2005PRL,Ishiwata2008,Ishiwata2010PRB,Kimura2012,Chai2014,Nakajima2016PRB,Zhai2017NatMat} with various structural types, including those at room temperature\cite{Kitagawa2010,Song2014,Chun2012}.
Among them, the largest $M$ switching ($\sim3\mu_B$/f.u.) by $E$ was obtained in Y-type hexaferrites~\cite{Chai2014,Zhai2017NatMat} at cryogenic temperatures, which is attributed to the simultaneous reversal of the ferroelectric and ferrimagnetic order parameters in a particular multiferroic phase, termed FE3 phase.
Moreover, this FE3 phase was found to emerge as a metastable state even at room temperature~\cite{Hirose2014,Nakajima2016PRB}.
Here we demonstrate that by choosing appropriate chemical composition and performing high-pressure oxygen annealing, the FE3 phase can be partially stabilized up to above room temperature.
This enables us to observe reversal of $M$ with considerable magnitude by $E$ field as well as the nearly-full reversal of $P$ by $H$ field in a single-component material near room temperature.
By using magnetic force microscopy (MFM) technique, magnetic domain switching by $E$ field is visualized.

Figure~\ref{ytype01}a shows the structural unit cell of the Y-type hexaferrite studied in the present work, Ba$_{0.8}$Sr$_{1.2}$Co$_2$Fe$_{12-x}$Al$_x$O$_{22}$ with $x$=0.9 (BSCFAO), which is composed of Fe$^{3+}$/Co$^{2+}$ and Fe$^{3+}$/Al$^{3+}$ ions in tetrahedral and octahedral oxygen coordinations, respectively, similarly to the other members of the material family.
It has been known~\cite{Kimura2012} that the magnetic structure in the hexaferrites is well described by ferrimagnetically-ordered spin-blocks with large ($\mathbf{S}^{\rm L}_{i}$) and small ($\mathbf{S}^{\rm S}_{i}$) net magnetizations alternately stacked along the $c$ axis.
As a result of complex magnetic interactions among the adjacent magnetic blocks, various magnetic structures have been identified~\cite{Ishiwata2010PRB,Nakajima2016PRB}.
These structures, such as commensurate phases FE3 and FE2' (Ref.~\onlinecite{Ishiwata2010PRB}), alternating longitudinal conical (ALC)~\cite{HakBongLee2011}, proper screw (PS), and collinear ferrimagnetic (FiM) phases (schematically illustrated in Fig.~\ref{ytype01}a and \ref{ytype01}b) are also observed in the present material.
The magnetic ground state reached via zero-field cooling was reported to be ALC for a hexaferrite with a similar composition~\cite{Chang2012PRB,Lee2012PRB}.
The FE3 phase is induced by $H$ field applied within the magnetic easy-plane, but preserved as a metastable state even after the field is removed~\cite{Hirose2014,Nakajima2016PRB}.
In the FE3 phase the magnetic moments of the $\mathbf{S}^\mathrm{L}$ and $\mathbf{S}^\mathrm{S}$ blocks form a double fan structure~\cite{Nakajima2016PRB}, lying in the $ab$ plane and a plane containing $c$-axis, respectively.
Spin-driven $P$ emerges within the $ab$ plane and perpendicular to the net $M$, due to the the inverse Dzyaloshinskii-Moriya mechanism~\cite{Ishiwata2008,Ishiwata2010PRB}.

The magnetic phases in BSCFAO have been investigated by the zero-field-cooled (ZFC) MFM, low-field-cooled magnetization, and neutron diffraction measurements as shown in Figs.~\ref{ytype01}c to \ref{ytype01}f.
The neutron diffraction measurements revealed a complex magnetic phase diagram with several co-existing magnetic orders (Fig.~\ref{ytype01}d and see Supplementary Information), which was determined by taking also the previous results into account~\cite{Ishiwata2010PRB,Nakajima2016PRB,Sagayama2009}.
Below $T_{\rm C1}$=450\,K, a magnetic peak with a commensurate wavevector of $q=3/2$ appears together with the onset of $M$ for $\mathbf{H}\perp{c}$, which indicates the coexistence of the FE3 and the collinear FiM phases.
At $T_{\rm C2}$=400\,K, the magnetic peaks with commensurate $q=3/4$ and incommensurate $q_{\rm IC}$ wavenumbers emerge, while $M$ for $\mathbf{H}\perp{c}$ decreases, indicating that the FiM phase is turned into the PS and FE2' phases.
Finally at $T_{\rm C3}$=300\,K, the PS order changes to the ALC phase as $M$ for $\mathbf{H}\parallel{c}$ shows a slight decrease.
Real-space MFM image of an $ac$ surface at room temperature indicates the phase separation between the strongly magnetic (large averaged-magnetization hosting) FE3/FE2' and weakly magnetic (little averaged-magnetization hosting) ALC/PS phases as shown in Fig.~\ref{ytype01}c (for details see the Supplementary Information).
A prominent feature of the present-composition compound, being distinct from the previous report~\cite{Nakajima2016PRB} on a similar Y-type hexaferrite, is the presence of stable FE3 phase among the zero-field-cooled states.

Magnetic state under $H$ applied within the $ab$-plane was investigated by magnetization and neutron-diffraction experiments.
In Fig.~\ref{ytype02}, $M$ and the neutron diffraction intensities corresponding to each of the co-existing phases are separably plotted.
Prior to the application of $H$, three phases coexist in the ZFC initial state in agreement with the temperature-dependent measurements.
At 100\,K (Fig.~\ref{ytype02}a), the ALC and FE2' phases disappear at $H$=2\,kOe and $H$=4\,kOe, respectively, while the FE3 phase takes over their places.
Once the single-phase state of FE3 is attained, it is fully preserved even when the $H$ field is removed or reversed.
On the contrary, at 250\,K (Fig.~\ref{ytype02}b) and 295\,K (Fig.~\ref{ytype02}c), both the ALC and FE2' phases reappear upon the reversal of the $H$ field.
At relatively high temperatures, thermal agitation is large enough to overcome the energy barriers between the competing phases with almost degenerated free energies, while not at low temperatures.
It is noted that magnetic anisotropy within the $ab$-plane is negligible at room temperature (Fig.~\ref{ytypeS_Manisotropy}), and hence the $M$-$H$ curve as well as the diffraction intensity are least affected by the anisotropy.

$H$-induced $P$ and $E$-controlled $M$ are shown in Fig.~\ref{ytype03}.  
Prior to the measurements, the single-domain ME state was attained by the application of $(+E_0,+H_0)$ poling fields in a crossed configuration ($\mathbf{E}\perp\mathbf{H};~\mathbf{E},\mathbf{H}\perp{c}$).
Below $T$=250\,K both $P$ and $M$ show anti-symmetric dependence on $H$ and $E$ fields, respectively, indicating that the $P$-$M$ coupling is conserved throughout the reversal of the fields.
Magnitude of the saturation value of the spin-driven polarization ($P^{\rm sat}$) is significantly larger than the earlier observations in other Y-type hexaferrites~\cite{Ishiwata2008,Chun2010,Chai2014,Nakajima2016PRB}, while comparable to TbMnO$_3$~\cite{Yamasaki2007} and the spin-driven component of BiFeO$_3$ which can be controlled by the field of more than $H$=100\,kOe~\cite{Tokunaga2015}.
Correspondingly, the magnetization change between $\pm{E}_{\rm max}$ fields, $\Delta{M}_{\rm E}$=5.5\,$\mu_B$/f.u., at $T$=100\,K is larger than in any former experiments performed at lower temperatures~\cite{Tokunaga2012NatPhys,Chai2014,Zhai2017NatMat}.
Even at 250\,K, a significant portion of $M$ can be reversed ($\Delta{M}_{\rm E} = 4.1\mu_B$/f.u.) by the $E$ field.
Near room temperature, symmetry of the $P$-$H$ and $M$-$E$ loops begins to change to a symmetric butterfly shape, indicating that the $P$-$M$ clamping is not fully preserved during the reversal.
Moreover, $P$-$H$ loops show a secondary hysteresis (indicated with black triangles), which is attributed to the re-emergence and disappearance of the PS and FE2' phases as shown in Fig.~\ref{ytype02}c. 

Importantly, the remanent $M$ of BSCFAO can be switched in a non-volatile manner between positive and negative values by $E$ field even at 250\,K, which is favourable for ME memory and spintronic applications.
Changes in the remanent $M$ for the first two $M$-$E$ loops are as large as $\Delta{M}_1$=3.9\,$\mu_B$/f.u. and $\Delta{M}_2$=3.0\,$\mu_B$/f.u. at 100\,K and $\Delta{M}_1$=2.5\,$\mu_B$/f.u. and $\Delta{M}_2$=1.5\,$\mu_B$/f.u. at 250\,K.
Correspondingly, the remanent $P$ is also switched between positive and negative values with ${P}^{\rm rem}$=95\,$\mu$C/m$^2$ at 250\,K.
As for the retention, $P$-$H$ loop exhibit good characteristics for the repeated reversal processes even at 295\,K (see Fig.~\ref{ytypeS03}).
However, $M$-$E$ loops are subject to deterioration at higher temperatures than 250\,K.
This decrease in the magnitude of the reversible $M$ is attributed to the weakened $P$-$M$ coupling as well as insufficient magnitude of the applicable $E$ field.

To further clarify the behaviour of the $P$-$M$ coupling, $M$ and $P$ reversal was investigated simultaneously in pulsed $E$ field experiments (Fig.~\ref{ytype04}).
Similarly to the quasi-static measurements, the single-domain FE3 state was initially prepared with $(+E_0,+H_0)$ poling fields ($\mathbf{E}\perp\mathbf{H};~\mathbf{E},\mathbf{H}\perp{c}$).
After the removal of the poling fields, triangular-shaped $E$-field pulse pairs were applied anti-parallel, then parallel with respect to the $E_0$ poling field (see Fig.~\ref{ytype04}a).
$M$ was measured before and after the pulses, while $P$ was measured during the same period as the $E$-field pulses were applied (see Methods for details).
The $\Delta{P}$-$E$ curve of magnetic origin at $T$=250\,K is displayed in Fig.~\ref{ytype04}d, where the partial reversal of the ferroelectric $P$ is attained by the pulsed $E$ field.

Figure~\ref{ytype04}e shows simultaneous reversal of the remanent $P$ and $M$ by four pairs of $E$-field pulses at 250\,K.
Upon the first negative $E$-field pulse, both the remanent $P$ and $M$ change from positive to negative, causing the magnetization change $\Delta{M}_{1}$=2.3\,$\mu_B$/f.u..
The $M$ is almost completely reversed at this point, and there is only a small change in $M$ for the second negative $E$-field pulse.
For the subsequent two positive $E$-field pulses, $P$ and $M$ were again reversed from negative to positive, with the change of $\Delta{M}_{2}$=1.8\,$\mu_B$/f.u..
Although the magnitudes of both $\Delta{P}$ and $\Delta{M}$ decrease as further pulses are applied, similarly to the quasi-static experiments, their parallel reduction demonstrates the strong $P$-$M$ clamping in this temperature range ($T\sim250$\,K).

Temperature dependence of the $P$ and $M$ switched by the first negative and positive $E$-field pulses, as defined as $\Delta{P}_1$, $\Delta{P}_2$, $\Delta{M}_1$, and $\Delta{M}_2$ (Figs.~\ref{ytype04}b, \ref{ytype04}c) respectively, are shown in Figs.~\ref{ytype04}f and \ref{ytype04}g.
Irrespective of the strength of the $P$-$M$ coupling, $P$ can be reversed by the $E$ field.
Magnitude of the reversed $\Delta{P}_1$ and $\Delta{P}_2$ slightly decreases as the temperature is increased, but remains finite at 300\,K, since the multiferroic FE3 phase is present in the whole temperature region shown here.
The initial value $M_0$, and the switched magnetizations $\Delta{M}_1$ and $\Delta{M}_2$ show similar temperature dependence with the $\Delta{P}$ up to 260\,K.
In contrast to the $\Delta{P}$, however, the $\Delta{M}$ exhibits more rapid decrease above 260\,K, and almost vanishes at 300\,K.
Therefore, the $E$-control over the $M$ is lost due to the weakened $P$-$M$ coupling rather than to the reduced volume fraction of the FE3 phase.

Using the real-space MFM imaging, we have investigated $E$-field induced motion of the magnetic domain walls (DW) at room temperature (Fig.~\ref{ytype05}).
The measurement was started from an initial state (0\,th in Fig.~\ref{ytype05}a), where poling $E$ and $H$ fields were once applied and then removed, and the evolution of the magnetic domain pattern of the same region was followed after several applications of the $E$ field (1st-2nd in Fig.~\ref{ytype05}a and 1st-4th Fig.~\ref{ytypeS_MFMsmall}).
As displayed in Fig.~\ref{ytype05}a, magnetic domain pattern clearly shows changes in response to the applied $E$ fields with different sign, which demonstrates that these are composite $P$-$M$ domain walls~\cite{Tokunaga2009NatMat}.
The most typical cases of domain dynamics are observed in regions R1 and R2.

At region R1 in Fig.~\ref{ytype05}a, the negatively magnetized region expands and shrinks due to the successive applications of $E$ field with alternate sign, which corresponds to DW propagation along the $c$ axis.
Figure~\ref{ytype05}b shows the MFM signals taken along the A-A' line, clearly demonstrating the DW motion along $c$ axis.
The process of $M$ switch is schematically illustrated in Fig.~\ref{ytype05}d, where one of the $M$ domains expands along the $c$ axis, so that the ME domain with $P$ parallel to $E$ expands. 
The small magnetic anisotropy within the $ab$ plane (see Fig.~\ref{ytypeS_Manisotropy}) suggests that in this boundary between the oppositely magnetized regions, local net $M$ is likely to rotate around the $c$ axis.

The region R2 in Figs.~\ref{ytype05}a and \ref{ytype05}c exemplifies a different process, where a positively magnetized domain is pushed in the image area from the upper side.
This behaviour is clearly illustrated in the line profile of Fig.~\ref{ytype05}c.
The process of $M$ switch (shown in Fig.~\ref{ytype05}e) is similar to the previous case, however, in this case the domains are separated by a DW, where local $M$ appears to form a cycloidal structure.
Apart from these successful examples, change in the ratio between the majority and minority magnetic domains is relatively small, pointing to the the decreased $P$-$M$ coupling at a relatively high temperature, e.g. room temperature.

In summary, we have demonstrated that the $M$ switching by $E$-field in BSCFAO is realized via the propagation of magnetic domain walls that respond to the $E$ field throughout the material.
At low temperatures $P$ and $M$ are tightly clamped and reversed simultaneously, while at high temperatures the $P$-$M$ coupling becomes weaker and the domain walls are deconfined.
To further improve the $E$-field induced $M$ reversal in Y-type hexaferrites, the confinement-deconfinement crossover of the domain walls should be pushed to higher temperature, while the co-existing ALC, PS, and FE2' phases have to be suppressed.

    \section*{Methods}
    
    \textbf{Single crystal growth, sample preparation and oxygen annealing procedures.}
    Single crystals of Y-type hexaferrite, Ba$_{0.8}$Sr$_{1.2}$Co$_2$Fe$_{12-x}$Al$_x$O$_{22}$ with $x$=0.9, were grown by the laser floating zone (LFZ) technique in 10\,atm oxygen atmosphere.
    First, SrCO$_3$, BaCO$_3$, Co$_3$O$_4$, Fe$_2$O$_3$ and Al$_2$O$_3$ were mixed in stochiometric amount and sintered in air at 1150\,$^{\circ}$C for 24\,h.
    Then the resulting product was pressed into rods and re-sintered for 14\,h in the same conditions.
    Y-type hexaferrite single crystals from earlier growths were used as seeds for the LFZ growth.
    The single crystal rods were oriented with a back-scattering Laue camera and cut into discs with the surfaces containing $c$-axis.
    To increase the resistivity of the samples for the ME as well as neutron diffraction measurements, the cut pieces were annealed in 10\,atm O$_2$ at 1000\,$^{\circ}$C for 100\,h in sealed quartz tubes, by adopting the technique described in Ref.~\onlinecite{Inaguma2001} (see the Supplementary Information and Fig.~\ref{ytypeS01a}).

	\textbf{Neutron diffraction measurements.}
	Neutron diffraction measurements were carried out at the triple-axis neutron spectrometer (PTAX) in the High Flux Isotope Reactor of Oak Ridge National Laboratory.
	Sliced and O$_2$-annealed single crystal of BSCFAO (approximately 25\,mm$^3$) were placed in a cryomagnet with $H$ applied along the $[010]$ axis, while $(h, 0, l)$ plane was set to be the scattering plane.
    
    \textbf{$P$-$H$, $M$-$E$ and $P$-$E$ measurements.}
    For each type of experiments, single crystals with the surface containing $c$-axis were coated with Au/Pt as electrodes, thus $E$ field was applied in the $ab$ plane while $H$ field was perpendicular to both the $E$ field and the $c$ axis ($\mathbf{E}\perp\mathbf{H};~\mathbf{E},\mathbf{H}\perp{c}$).
    $H$ field dependence of the polarization was measured in a PPMS (Quantum Design) with an electrometer (Keithley 6517A) by monitoring the displacement current as the $H$ field was swept with 100\,Oe/s continuously between $\pm$5\,kOe for 11-21\,cycles, depending on the signal to noise ratio.
    Current peaks around 0\,Oe did not show degradation, therefore the $P$-$H$ curves were obtained by integrating the current after averaging.
    Magnetization measurement under $E$ field was carried out in an MPMS-XL (Quantum Design), while the electrometer (Keithley 6517A) was used as a voltage source.
    The thickness, surface area, and mass of the sample were 70\,$\mu$m, 1.64\,mm$^2$, and 0.71\,mg, respectively.
    Pulsed $E$ field measurements were performed with a ferroelectric tester (Radiant Inc., Precision Premiere II) equipped with 500\,V option.
    The ferroelectric polarization of magnetic origin was measured by the Positive-Up-Negative-Down (PUND) technique.
	At low temperatures, triangular-shaped $E$-field pulses with 7\,MV/m in amplitude and 50\,ms in duration were applied.
	Above 280\,K, however, the pulse duration was reduced to 1\,ms due to the lower resistivity.
	
	\textbf{Magnetic force microscopy measurements.}
	Magnetic force microscopy measurements were carried out with a commercially available scanning probe microscope (MFP-3D, Asylum Research) using Co-coated cantilever (MFMR-10, Nano World).
	For $E$-field-dependent measurements, samples were poled to a single domain ME state using +3\,MV/m and +4\,kOe poling fields in $\mathbf{E}\perp\mathbf{H};~\mathbf{E},\mathbf{H}\perp{c}$ configuration in a PPMS (for sample preparation see Supplementary Information).
	Static $E$ field (+3\,MV/m or $-$3\,MV/m) was applied to manipulate magnetic domains using a Keithley 6517A electrometer, and the MFM images were taken after the $E$ field was switched off.
	The sign and magnitude of  MFM phase shift,  $\Delta\varphi$, roughly
correspond to those of the magnetization perpendicular to the plane.

\textbf{Acknowledgements} This research used resources at the High Flux Isotope Reactor, a DOE Office of Science User Facility operated by the Oak Ridge National Laboratory. This research was supported in part by the U.S.-Japan Cooperative Program on Neutron Scattering. Structural unit cell of the Y-type hexaferrite crystal was illustrated using the software \texttt{VESTA}\cite{Momma2011}. The authors are grateful for the technical assistance provided by Maximilian A. Hirschberger.

\textbf{Author Contributions} V.K., T.N., M.M., K.K., T.A. performed the measurements and analyzed the data; V.K., Y.K., A.K., Y. Tokunaga prepared the single crystal samples; J.T. investigated the O$_2$ annealing; V.K., F.K. took the MFM images; V.K., Y. Taguchi wrote the manuscript; Y. Taguchi and Y. Tokura conceived the project; all the authors contributed to the discussion of the results.


\pagestyle{empty}

    \begin{center}
    \begin{figure*}[t!]
 
    \includegraphics[width=16.5truecm]{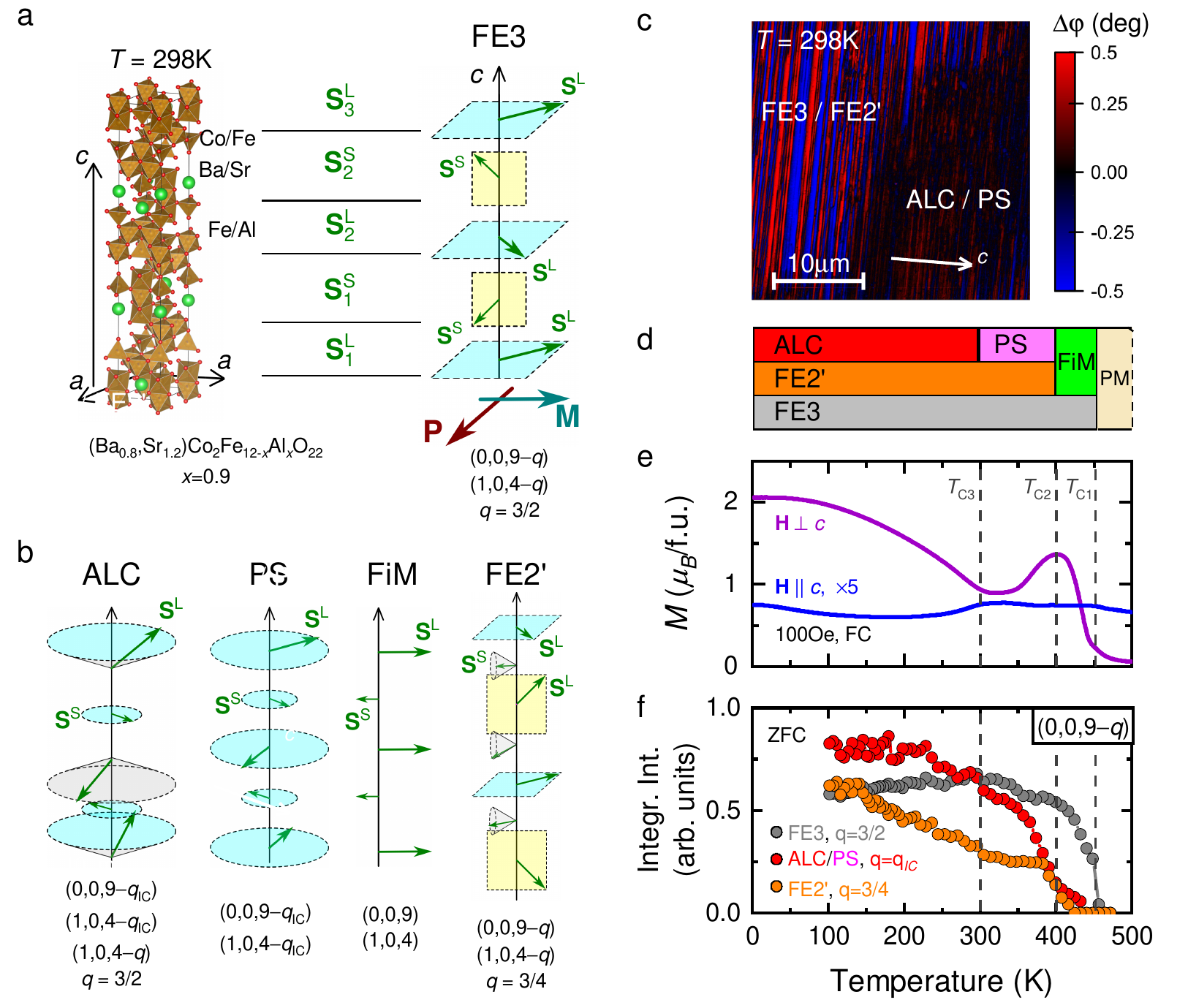}
    \caption{\textbf{ Structural and magnetic properties of Ba$_{0.8}$Sr$_{1.2}$Co$_2$Fe$_{12-x}$Al$_x$O$_{22}$ ($x$=0.9, BSCFAO).}
    \textbf{a}-\textbf{b}, Schematic of the structural and magnetic unit cells, the latter of which is composed of alternately stacked spin-blocks with large ($\mathbf{S}^\mathrm{L}$) and small ($\mathbf{S}^\mathrm{S}$) magnetic moments. Alternate longitudinal conical (ALC), proper screw (PS), ferrimagnetic (FiM) orders as well as the multiferroic FE3 and FE2' phases are illustrated in terms of the spin-blocks. The FE3 phase can be viewed as a double fan structure, where the $\mathbf{S}^\mathrm{L}$ and $\mathbf{S}^\mathrm{S}$ spins are staggered in the $ab$ plane and the plane including $c$-axis, respectively, and the net magnetization  ($\mathbf{M}$) and net polarization ($\mathbf{P}$) are perpendicular to each other and also to the $c$ axis. Respective magnetic phases are identified by the characteristic magnetic reflections indicated below each of the magnetic structure.
    \textbf{c}, Real-space magnetic force microscopy (MFM) image of a BSCFAO sample with an $ac$ surface after zero field cooling. The striped and dark regions correspond to the magnetoelectric FE3/FE2' and the incommensurate ALC/PS phases with typical dimensions of 5-30\,$\mu$m with high and low contrast of MFM phase signals, respectively, indicating the phase separation. The magnetic domain within the FE3/FE2' region has 200-300\,nm in thickness along the $c$ axis and 10-20\,$\mu$m in length along the $ab$ plane.
    \textbf{d}, Zero-field-cooled magnetic phase diagram presented as a chart diagram, showing coexistence of the magnetic phases. Horizontal axis is common for panels d-f.
    \textbf{e}, Temperature dependence of the field-cooled $M$ in $H$=100\,Oe for $\mathbf{H}\parallel{c}$ and $\mathbf{H}\perp{c}$.
    \textbf{f}, The integrated intensities of selected neutron diffraction peaks representing the different magnetic phases in the zero-field-cooled measurements are plotted against temperature.}
    \label{ytype01}
    \end{figure*}
    \end{center}

    \begin{center}
    \begin{figure*}[t]

    \includegraphics[width=16truecm]{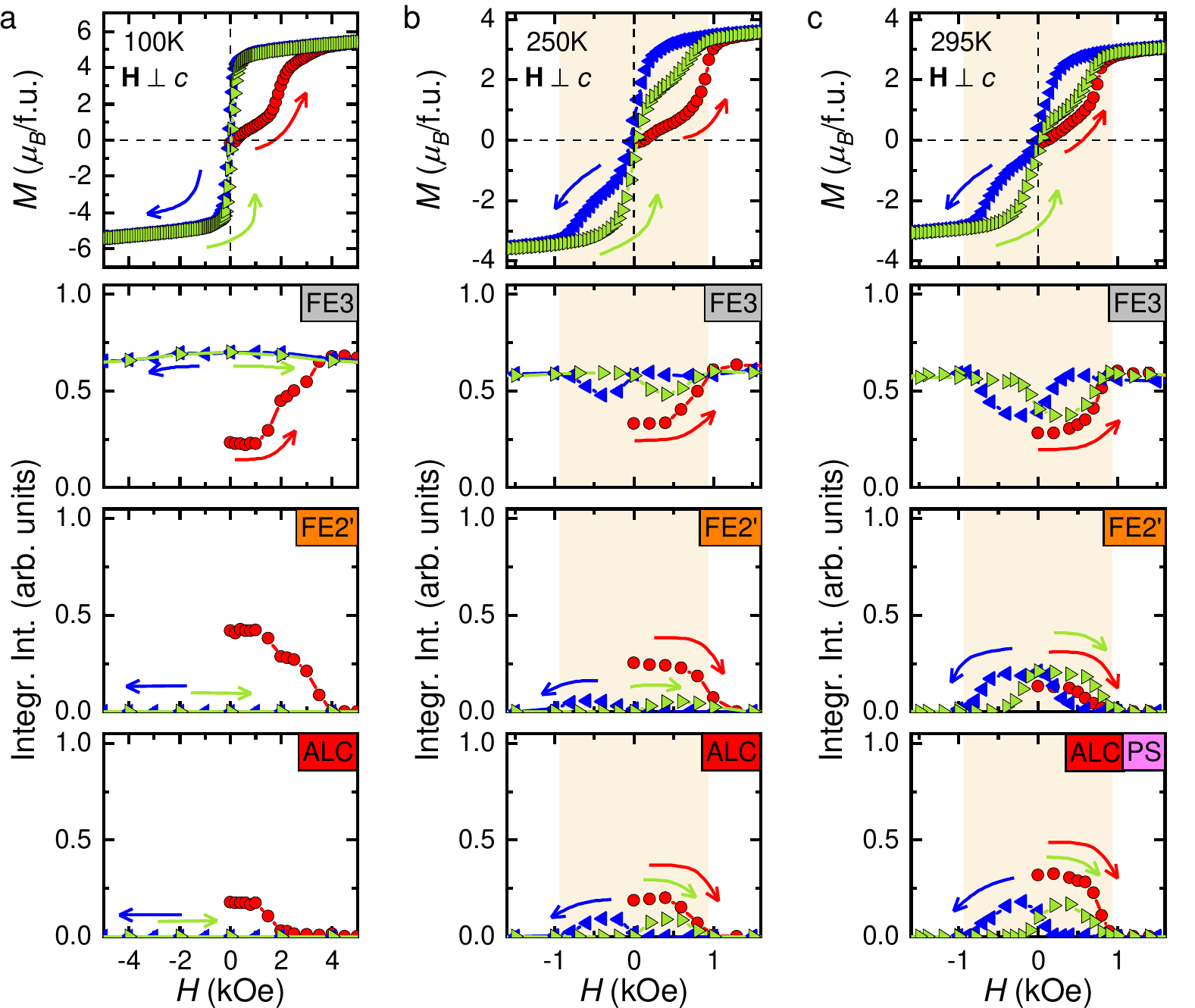}
    \caption{\textbf{ Relationship between the magnetization and the magnetic phases in magnetic field.}
    Magnetic-field ($H$) dependence of magnetization and integrated intensity of neutron diffraction peaks for relevant phases at \textbf{a}, 100\,K, \textbf{b}, 250\,K and \textbf{c}, 295\,K.
    The measurements were started from a zero-field-cooled state, then $H$ field was applied perpendicular to the $c$ axis.
    The FE3, FE2', and ALC/PS phases are represented by the neutron diffraction peaks $(0,0,9-q)$ with $q=3/2$, $q=3/4$ commensurate, and $q_{\rm IC}$ incommensurate wavenumbers, respectively.
    In all the panels, red symbols indicate the initial $H$-increasing process, while blue and green symbols denote field-decreasing and second increasing runs, respectively.
    The ALC phase can be identified in the $M$-$H$ measurements as exhibiting low $M$ and low $\frac{dM}{dH}$ (not shown) appearing at the low $H$ region near the origin, while the FE3 phase has high $M$, showing distinction from the ALC phase.
    At $T$=100\,K (\textbf{a}), when magnetic field of 5\,kOe is applied, the FE2' and ALC phases disappear, leaving FE3 the only phase. Once the FE3 phase is stabilized, it is preserved throughout the subsequent reversal processes of the $H$ field. 
    At $T$=250\,K (\textbf{b}), the FE2' and ALC phases re-emerge when the $H$ field is reversed. The fraction of the re-appearing phases becomes largest not at zero field, but at $\pm$500\,Oe.
    Finally at $T$=295\,K (\textbf{c}), ratio of the re-appearing phases becomes larger, but the FE3 phase is partially preserved.}
    \label{ytype02}
    \end{figure*}
    \end{center}

    \begin{center}
    \begin{figure}[t]

    \includegraphics[width=8.5truecm]{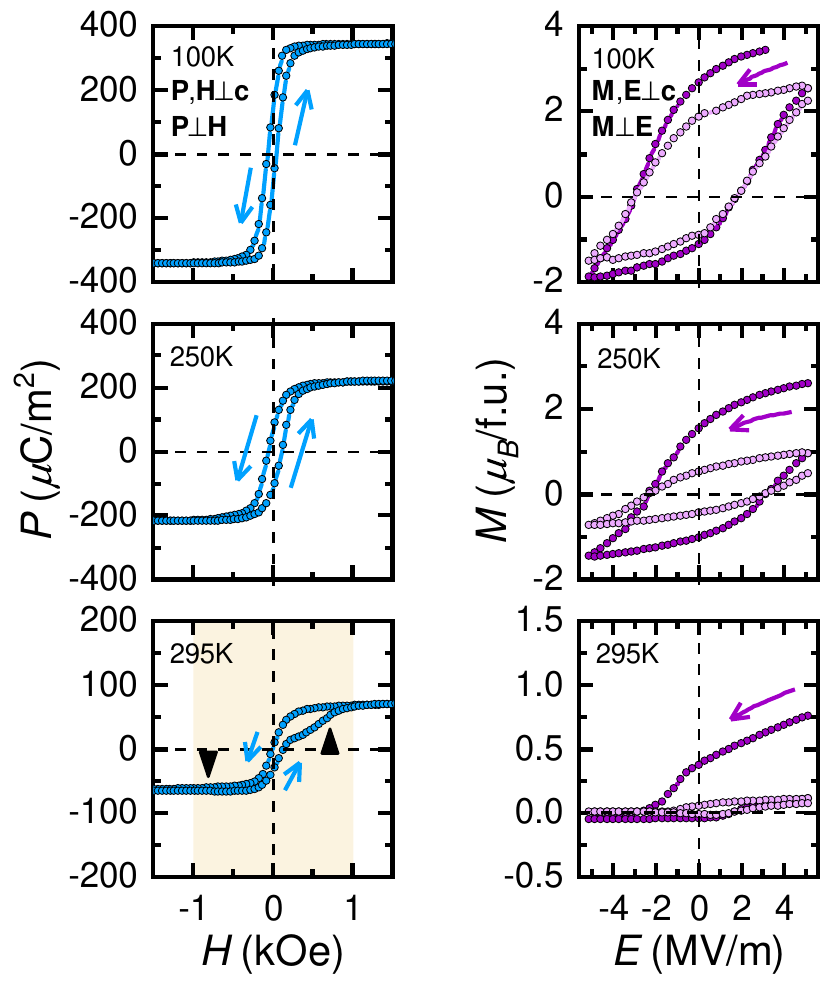}
    \caption{\textbf{ Cross-control of polarization and magnetization.}
    Prior to the measurements, a single-domain ME state was prepared by $(+E_0,+H_0)$ poling fields applied in $\mathbf{E}\perp\mathbf{H};~\mathbf{E},\mathbf{H}\perp{c}$-axis configuration.
    The $P$-$H$ and $M$-$E$ measurements were performed in the absence of $E$ and $H$ fields, respectively.
    At low temperatures, both $P$-$H$ and $M$-$E$ curves exhibit anti-symmetric shape with respect to the fields, indicating that the $P\times{M}$ is conserved due to the strong $P$-$M$ coupling.
    Around room temperature, components with symmetric field-dependence are mixed in the $P$-$H$ and $M$-$E$ loops, implying that the $P\times{M}$ is not fully preserved anymore. The region where the ALC and PS phases re-emerge is highlighted by light shading in the $P$-$H$ data at 295\,K, while the secondary hysteresis of the $P$-$H$ loops are indicated by black triangles.}
    \label{ytype03}
    \end{figure}
    \end{center}

    \begin{center}
    \begin{figure*}[t]

    \includegraphics[width=17truecm]{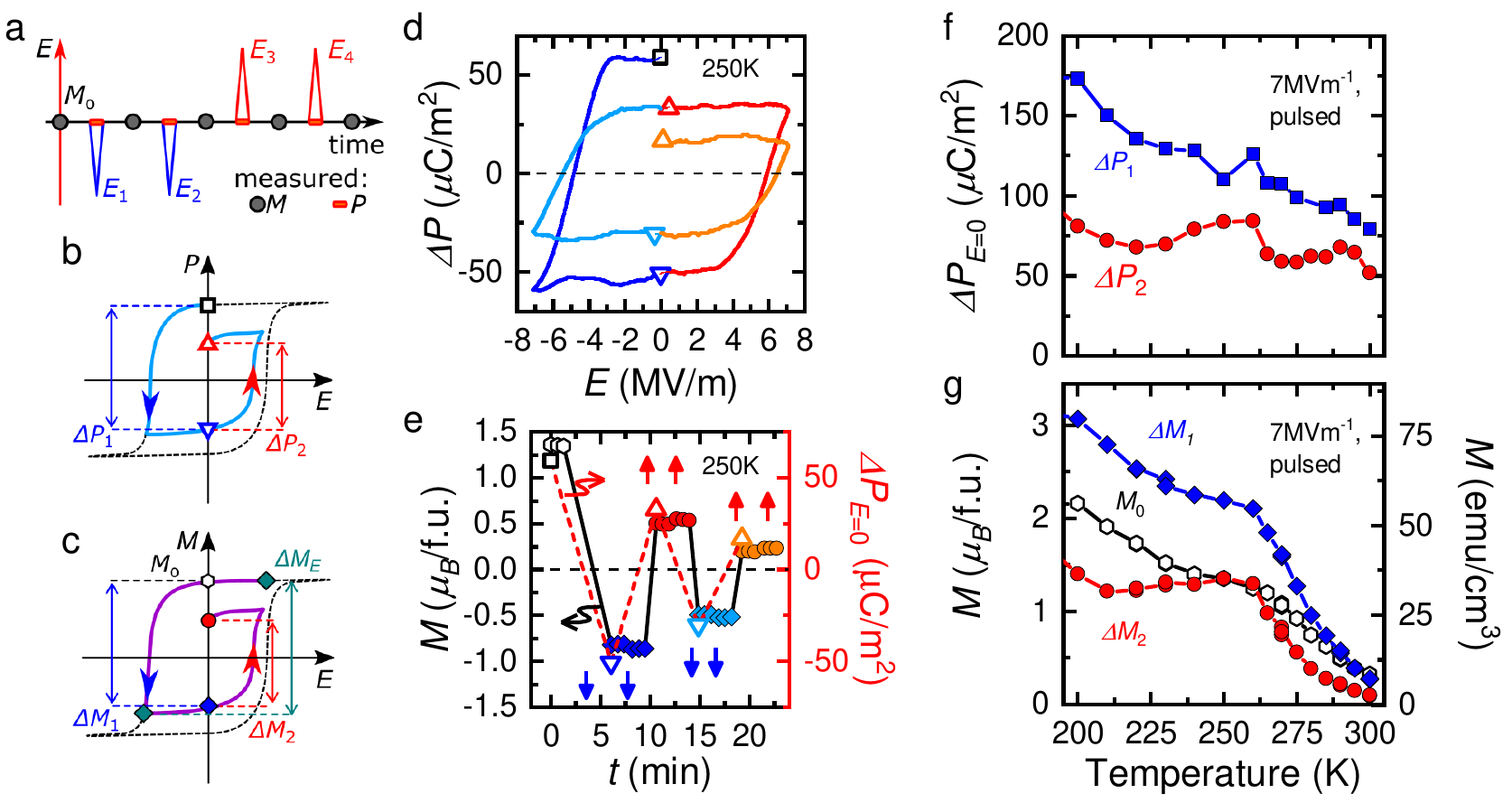}
    \caption{\textbf{ Magnetization reversal by pulsed electric field.} 
    \textbf{a}, Schematic of the experimental procedure. Measurements were started from a single domain ME state achieved by ($+E_0$,$+H_0$) poling in a $[\mathbf{E}\perp\mathbf{H};~\mathbf{E},\mathbf{H}\perp{c}]$ configuration. Triangular-shaped $E$-field pulses were applied to the sample. Magnetization was measured before and after the pulses, while polarization at the same time. Duration ($\tau$) of the pulses was 50\,ms and 1\,ms for $T<280$\,K and $T\geqslant280$\,K, respectively.
    \textbf{b}, Schematic illustration of the definitions for the $\Delta{P}_{1}$ and $\Delta{P}_{2}$ polarization changes. A minor loop without saturation of $P$ is represented by red line.
    \textbf{c}, Schematic $M$-$E$ loop showing the definitions for $M_0$, $\Delta{M}_{1}$, $\Delta{M}_{2}$ and $\Delta{M}_{\rm E}$.
    A minor loop is illustrated by a green curve,  $M_0$ is the initial remanent magnetization after poling, and $\Delta{M}_{1}$, $\Delta{M}_{2}$ are the magnetization changes after $-E$ and $+E$ pulses. The $\Delta{M}_{\rm E}$ is defined only for quasi-static experiments (Fig.~\ref{ytypeS06}) as the magnetization difference between $+E_0$ and $-E_0$ fields.
    \textbf{d}, The polarization change of magnetic origin determined from the pulsed $E$-field experiments at 250\,K after $(+E_0,+H_0)$ poling.
    \textbf{e}, Changes in the remanent $M$ and $P$ induced by the $E$-field pulses at 250\,K. Blue downward and red upward arrows indicate the sign of the $E$ pulses.
    Although both $M$ and $\Delta{P}_{E=0}$ decrease due to the insufficient $E$-field strength for complete reversal, their parallel change demonstrates their strong coupling.
    Temperature dependence of \textbf{f}, $\Delta{P}_{1}$ and $\Delta{P}_{2}$, and \textbf{g}, $M_0$, $\Delta{M}_{1}$ and $\Delta{M}_{2}$ switched by the first and second $E$-field pulses. Up to 260\,K, the $E$-field induced $\Delta{P}$ and $\Delta{M}$ exhibit similar temperature variation. Upon approaching $T$=295\,K, temperature dependence of $\Delta{P}$ and $\Delta{M}$ shows clear departure, indicating that the $P$-$M$ coupling starts to get weaker. From a technological viewpoint, the magnetization changes are shown also in emu/cm$^3$ unit.}
    \label{ytype04}
    \end{figure*}
    \end{center}

    \begin{center}
    \begin{figure*}[t]

    \includegraphics[width=17truecm]{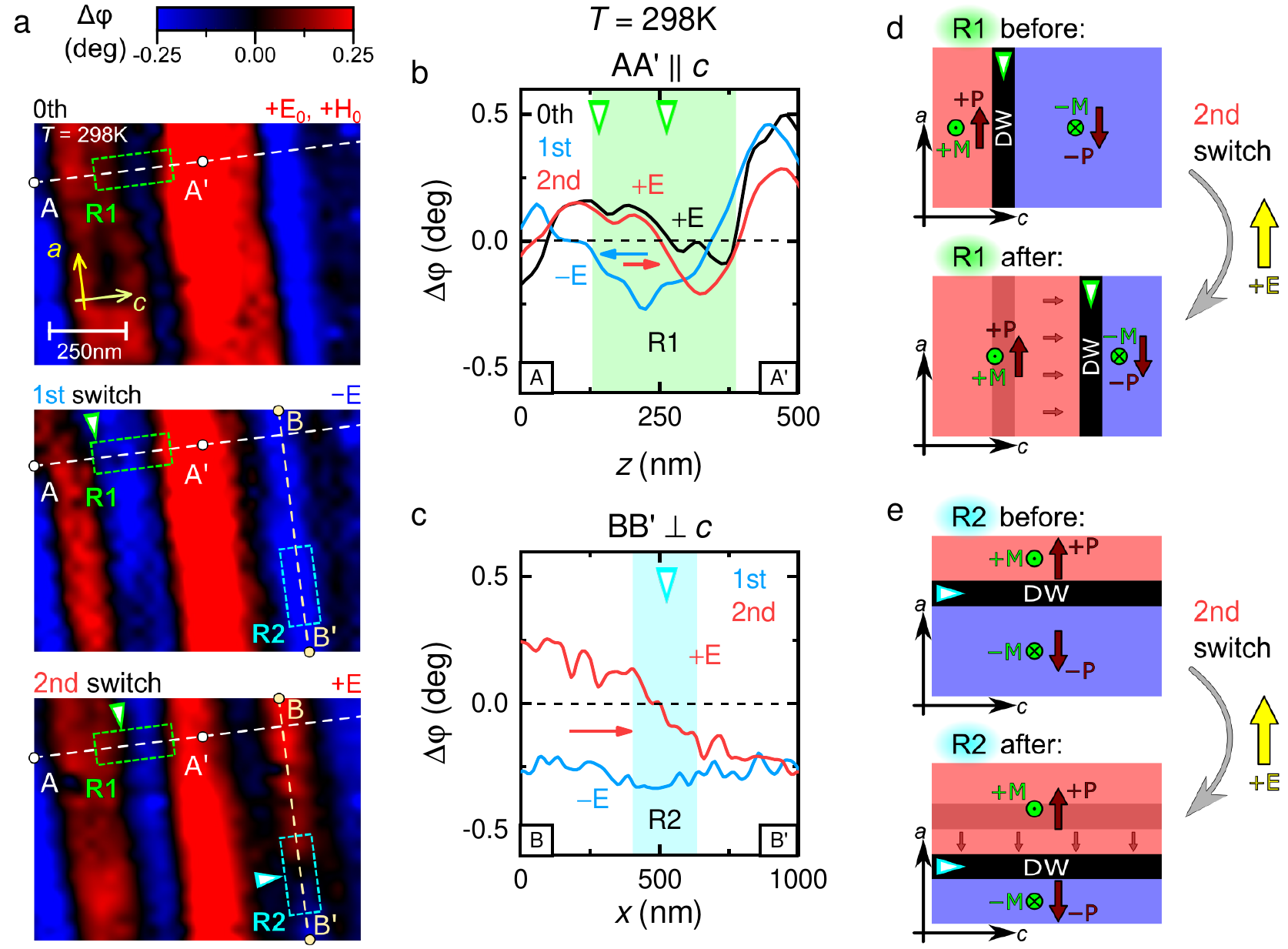}
    
    \caption{\textbf{ Real-space magnetic force microscopy (MFM) images of the $E$-field induced changes in the magnetic domains at room temperature.}
    The MFM images were taken on the same 10$\times$10\,$\mu$m$^2$ region of a BSCFAO crystal with an $ac$ face (see Figs.~\ref{ytypeS01c}, \ref{ytypeS_MFMlarge}, and \ref{ytypeS_MFMsmall}).
    Prior to the MFM measurements, the sample was poled to a single domain ME state using ($+E_0$,$+H_0$) poling fields in a $\mathbf{E}\perp\mathbf{H};~\mathbf{E},\mathbf{H}\perp{c}$ configuration.
    Panel \textbf{a} shows the changes in the magnetic domain pattern caused by two successive applications of the $E$ field with different signs (the initial state is labeled as the 0\,th).
    The images include small regions, R1 and R2, where two representative cases of DW motion are observed.
    Around R1, the negatively magnetized domain (denoted with blue colour, MFM phase shift $\Delta\varphi<0$) expands and shrinks along the $c$ axis upon the first and second applications of $E$-field, respectively.
    On the other hand, around R2, a positively magnetized domain (denoted with red colour, $\Delta\varphi>0$) is pushed into the view area from the upper side along the $ab$ plane.  
    These two cases are further displayed as line profiles of the MFM phase shift ($\Delta\varphi$) data along the \textbf{b}, A-A' and \textbf{c}, B-B' lines.
    Panels \textbf{d} and \textbf{e} show the schematic illustration of these two cases of domain wall motions for the second $E$-field switch, respectively.}
    \label{ytype05}
    \end{figure*}
	\end{center}

\renewcommand{\thefigure}{S\arabic{figure}}
\renewcommand{\theequation}{S\arabic{equation}}
\renewcommand{\thetable}{S\arabic{table}}
\setcounter{figure}{0}
\cleardoublepage

\begin{center}
\textbf{Supplementary Material}
\end{center}

\textbf{Effect of oxygen annealing on resistivity}

Sliced samples were annealed in 10\,atm O$_2$ at 1000\,$^{\circ}$C for 100\,h in sealed quartz tubes, similarly to Ref.~\onlinecite{Hirose2015JACS}, by adopting the technique reported in Ref.~\onlinecite{Inaguma2001} using Ag$_2$O as oxygen source.
Figure~\ref{ytypeS01a} presents the electrical resistivity of BSCFAO measured by a two-probe method before (black curve) and after (red curve) the O$_2$ annealing procedure under constant voltage and 2\,K/min temperature sweep rate.
For the sake of comparison, the resistivity determined from the $I$-$V$ curves simultaneously obtained during the $M$-$E$ measurements at fixed temperatures are shown by orange coloured dots.
Resistivity of BSCFAO shows semiconductor-like temperature dependence with an activation energy of 0.5\,eV prior to the O$_2$ heat-treatment.
After the the O$_2$ annealing, the activation energy is increased to 1\,eV and the resistivity is enhanced by 5-6 orders of magnitude.
The as-grown crystals have oxygen deficiencies, which indicates that the material has Fe$^{2+}$ besides the Fe$^{3+}$ ions, thereby leading to higher conductivity.
The high-pressure annealing reduces the oxygen deficiencies and increases the resistivity. 

\vspace*{18pt}
\textbf{Sample preparation procedures for magnetoelectric (ME) and magnetic force microscopy (MFM) mearurements}

Sample preparation processes for the magnetoelectric measurements are illustrated in a step-by-step manner in Fig.~\ref{ytypeS01b}a.
Both for $P$-$H$ and $M$-$E$ measurements, single crystals were sliced so that the $c$-axis is parallel to the surfaces, which were coated with Au/Pt electrodes, thus $E$ field was applied within the $ab$ plane while the $H$ field was perpendicular to both the $E$ field and the $c$ axis ($\mathbf{E}\perp\mathbf{H}$; $\mathbf{E},\mathbf{H}\perp{c}$).
In case of $M$-$E$ measurements, coverage of the surfaces by the electrodes was crucial, as the $M$ of the uncovered parts cannot be reversed.
Therefore, the edges of the samples were cut off in order to ensure the complete coverage.
The cut samples were placed on sapphire plates, where the bottom electrode was prepared with heat-treatment silver paste (Dupont 7095), while the top electrode was connected with a gold wire and silver paste (Dupont 4922N).
A photograph of the sample used in the $M$-$E$ measurements is shown in Fig.~\ref{ytypeS01b}b.
$H$ field was applied perpendicular to the $E$ field and the $c$ axis along the lateral direction.
This piece had 3\,mm lateral dimension, 1.64\,mm$^2$ surface area, 70\,$\mu$m thickness, and 0.71\,mg mass.

Photos of the sample for the $E$-field dependent MFM measurements are presented in Fig.~\ref{ytypeS01c}.
The sample was prepared with the following method.
First, high-quality surface needed for the experiment was obtained by mechano-chemical polishing with 0.050\,$\mu$m silica suspension on a BSCFAO sample with $ac$ surface.
Grooves were cut parallel to the $c$ axis in the surface with use of a wire saw (WS22, K.D. Unipress), and silver paste was used to form the electrodes and to fix gold wires.
Distance between the electrodes in the present experiment is $\sim$80\,$\mu$m.
During the poling procedure, the $E$=3\,MV/m and $H$=4\,kOe fields were applied within and perpendicular to the surface, respectively in the $\mathbf{E}\perp\mathbf{H};~\mathbf{E},\mathbf{H}\perp{c}$ configuration, as shown in Fig.~\ref{ytypeS01c}a. 

\vspace*{18pt}
\textbf{Neutron diffraction measurements}

Neutron diffraction measurements were performed on a thin plate (thickness $\sim$1\,mm) of O$_2$-annealed single crystals with 25\,mm$^3$ volume.
Neutron diffraction profiles along $(0,0,l)$ direction at selected temperatures and the corresponding wavenumbers of the FE3, FE2', ALC and PS magnetic orders are presented in Fig.~\ref{ytypeS02}a and \ref{ytypeS02}b.
In the paramagnetic phase, at $T$=500\,K, only the two nuclear reflections are present at $l$=6 and $l$=9.
At $T$=450\,K, a strong magnetic peak appears at $l$=7.5, which identifies the FE3 phase with commensurate $q$=3/2 wavenumber ($l=9-q$).
At around $T$=400\,K, magnetic reflections with double-peak structure appear, corresponding to the commensurate $q$=3/4 and incommensurate $q_{\rm IC}$ wavenumbers.
These magnetic reflections represent the FE2' and the ALC/PS phases, respectively.
The $q_{\rm IC}$ wavenumber shows a weak and non-monotonous temperature dependence, while the wavenumbers of FE3 and the FE2' phases remain fixed to commensurate positions (Fig.~\ref{ytypeS02}b).

In Fig.~\ref{ytypeS02}e, temperature dependence of the neutron diffraction integrated intensities is displayed.
The spin structures of the FE3 and FE2’ phases are regarded as being composed of ferrimagnetic and cycloidal components that are parallel and perpendicular to the net magnetization direction, respectively, as depicted in Fig.~\ref{ytype01} of Ref.~\onlinecite{Nakajima2016PRB}.
Therefore, the neutron intensities at $(0,0,9)$ and $(1,0,4)$ contain nuclear scattering and magnetic scattering from the ferrimagnetic component which has the same periodicity as the chemical lattice.
As the temperature is decreased from 400\,K to 300\,K, the in-plane magnetization as well as the intensities at $(0,0,9)$ and $(1,0,4)$ are decreased, indicating that the $q$=0 ferrimagnetic components are reduced.
On the other hand, the magnetic reflections corresponding to the wave vectors of $(0,0,3/2)$, $(0,0,3/4)$ and $(0,0,q_{\rm IC})$ remain unchanged or rather increase in this temperature range, suggesting that the reduction of the intensities at $(0,0,9)$ and $(1,0,4)$ are ascribed solely to the disappearance of the FiM order.
The gradual increase below 300\,K can be explained by an increase of volume fraction of the FE2’ phase.
In this series of Y-type hexaferrites, many previous studies have reported that PS or (alternating) longitudinal conical phase appears from the FiM phase as the temperature is lowered~\cite{HakBongLee2011,Lee2012PRB}.
Although it could be possible to consider that the FiM phase still survives at low temperatures, the volume fraction of the residual FiM phase is likely to be very small.
Therefore, the FiM phase can simply be viewed as a collinear parent phase, from which PS/ALC or FE2’ emerges by spin canting or spin rotating at lower temperatures, and hence not relevant for the discussion on the observed magnetoelectric properties.

In Fig.~\ref{ytypeS02}f, half width at half maximum (HWHM) of each reflection is plotted against temperature.
The $(0,0,9)$ nuclear peak serves as resolution limit for the HWHM of the magnetic peaks.
As shown in Fig.~\ref{ytypeS02}f, the magnetic satellite peaks have larger HWHM than the nuclear peak, which implies the presence of magnetic domains.
Moreover, respective magnetic satellite peaks have different HWHMs, indicating that the corresponding domain sizes may be different.
If a third phase with multiple-$q$ existed instead of phase separation, these correlation lengths should have been the same value.
Therefore, phase coexistence of single-$q$ magnetic phases is more likely.

\vspace*{18pt}
\textbf{MFM measurements after zero-field-cooling}

Zero-field MFM measurements were performed at room temperature on an O$_2$ annealed sample with mechano-chemically polished $ac$ surface.
This sample is different from the one used for the $E$-field dependent MFM experiments.
Figures~1c and \ref{ytypeS_MFMhdep} demonstrate the spatial separation of the coexisting magnetic phases in BSCFAO.
The striped regions correspond to the phase with large $M$ perpendicular to the surface of the sample (namely the FE3/FE2'), as the modulation of the MFM phase shift is large.
On the other hand, the regions with low MFM signal are the incommensurate magnetic phases with small magnetization, namely the ALC/PS phases.

The magnetic phases were further investigated by changing the distance between the cantilever and the surface of the sample ($\Delta{h}$, representing the difference in the cantilever height between the topography imaging and the MFM imaging).
Figure~\ref{ytypeS_MFMhdep}b and \ref{ytypeS_MFMhdep}c show the change in the MFM signal for different values of the $\Delta{h}$.
By increasing the distance from the surface, magnitude of the MFM signal decreases, however, this change is larger at those regions where the modulation is originally higher (Figs.~\ref{ytypeS_MFMhdep}b and \ref{ytypeS_MFMhdep}c).
This suggests that in the areas with low modulation, an FE3/FE2' phase is buried beneath the the ALC/PS phase, as schematically illustrated in Figs.~\ref{ytypeS_MFMhdep}d and \ref{ytypeS_MFMhdep}e.

In the present setup, the spatial resolution of the measurement is $\sim$30\,nm, while the observed stripes have 200-300\,nm in width along the $c$ axis and 10-20\,$\mu$m in length along the $ab$ plane.
Therefore it is justified to assume that the striped pattern observed by the MFM measurement represents the magnetic domain pattern, while precise determination of the thickness along the $c$ axis or the internal structures of the domain walls (DW) using MFM experiments alone is difficult.

\vspace*{18pt}
\textbf{Magnetic anisotropy}

Information on the magnetic anisotropy is useful for the understanding of the magnetic domain and domain wall structures.
The FE3 phase of BSCFAO has easy-plane magnetic anisotropy perpendicular to the crystallographic $c$ axis at $T$=300\,K (see Fig.~\ref{ytypeS_Manisotropy}a).
When the magnetic field is rotated within the easy-plane ($ab$-plane), a six-fold modulation is expected in the angular dependence of the $M$ due to the trigonal crystal symmetry.
As demonstrated in Figs.~\ref{ytypeS_Manisotropy}b and \ref{ytypeS_Manisotropy}c, the magnetic anisotropy within the $ab$ plane is negligibly small $\delta{M}(\vartheta)=M(\vartheta) - M_{\rm ave}$, where $M_{\rm ave}$ is the averaged magnitude of $M$ over 360$^{\circ}$.
The six-fold modulation is as small as 0.1$\%$ of $M_{\rm ave}$ at 300\,K and $H$=1\,kOe, where the system is in the FE3 phase, and remains less than 0.7$\%$ at $T$=5\,K.
Therefore at room temperature, the net magnetic moment of the FE3 phase confined in the $ab$ plane take arbitrary direction as a result of the small anisotropy within the $ab$ plane.

\vspace*{18pt}
\textbf{Magnetoelectric poling procedure}

Prior to the $P$-$H$ and $M$-$E$ measurements, a single-domain FE3 state was prepared by isothermal ME poling with fields applied in the $\mathbf{E}\perp\mathbf{H}$; $\mathbf{E},\mathbf{H}\perp{c}$ configuration.
In this poling procedure, the single-domain FE3 state is attained by decreasing the $H$ field, i.e. approaching from the high-field FiM phase, in the presence of $E_0$ field.
In case of $P$-$H$ measurements, $E_0=0.4$\,MV/m and $H_0=50$\,kOe poling fields were applied at $T$=250\,K, then the magnetic field was reduced to 5\,kOe.
After turning off the $E$ field, temperature was swept to the measurement temperature in the presence of the 5\,kOe field.
In case of $M$-$E$ measurements, the $E_0=5$\,MV/m and $H_0=10$\,kOe poling fields were applied at the same temperature as the measurement was performed.

\vspace*{18pt}
\textbf{Measurement of $H$-dependent polarization}

The spin-induced displacement current for repeated cycles at 295\,K, used for the calculation of the $P$-$H$ curve, is shown in Fig.~\ref{ytypeS03}.
$P$-$H$ measurements were started from a single domain ME state prepared by the isothermal poling procedure described before.
The displacement current was monitored as the $H$ field was swept between $\pm$5\,kOe with a rate of 100\,Oe/s.
When the $H$ is reversed from positive to negative, a large negative current peak was observed, indicating the reversal of the $P$ (Fig.~\ref{ytypeS03}a).
For the $-H\rightarrow{+H}$ sweep, current pulse with double-peak structure was detected, which is associated with the reversal of $P$ and the secondary hysteresis connected to the re-appearance of the FE3 phase.
The displacement current was averaged for 21 cycles (orange curve) to obtain a single $I$-$H$ loop (blue curve) in Fig.~\ref{ytypeS03}b, which was integrated to obtain $P$, and the resultant $P$-$H$ curve is presented in Fig.~\ref{ytype03}.

\vspace*{18pt}
\textbf{Temperature dependence of $P$-$H$ and $M$-$E$ properties in quasi-static measurement}
Figure~\ref{ytypeS06} shows the temperature dependence of the direct and the converse ME effects obtained from the quasi-static measurements.
The saturation ($P^{\rm sat}$) and remanent ($P^{\rm rem}$) values of the spin-induced $P$, determined from $P$-$H$ loops (see Fig.~\ref{ytype03} and Fig.~\ref{ytypeS03}) are presented in Fig.~\ref{ytypeS06}a, while the initial $M_0$, the $\Delta{M}_1$, $\Delta{M}_2$ and $\Delta{M}_{\rm E}$ changes, obtained from the $M$-$E$ measurements (see Fig.~\ref{ytype04}c for the definitions), are shown in Fig.~\ref{ytypeS06}b.
The $P$-$H$ characteristics have different temperature dependence compared to the $M$-$E$ characteristics, as discussed in the main text.
$\Delta{P}_1$ and $\Delta{P}_2$ show gradual decrease towards 300\,K, in contrast to $P^{\rm rem}$, which completely disappears at 305\,K.
Magnitudes of the spin-induced $P^{\rm sat}$ and $P^{\rm rem}$,  as well as $\Delta{P}_1$ and $\Delta{P}_2$ are related to the volume fraction of the FE3 phase.
However, $P^{\rm rem}$ and $P^{\rm sat}$ polarization values are those reversed by $H$-field, and hence should be affected by the $P$-$M$ coupling, while $\Delta{P}_1$ and $\Delta{P}_2$ are not since they are driven directly by pulsed $E$-field, which explains the difference in the temperature dependence.
Characteristic values associated with the $M$-$E$ loops exhibit analogous temperature dependence between the quasi-static and pulsed $E$ field measurements.
Moreover, the initial $M_{0}$ and the change $\Delta{M}_{\rm E}$ follow similar temperature dependence to the $P^{\rm rem}$ and $P^{\rm sat}$, respectively.

\vspace*{18pt}
\textbf{$E$-field induced change in magnetic domains as revealed by MFM measurements}

MFM measurements before and after the first application of $E$ field of $-$3\,MV/m for the 10$\times$10\,$\mu$m$^2$ area (shown in Fig.~\ref{ytypeS01c}) are presented in Figs.~\ref{ytypeS_MFMlarge}a and \ref{ytypeS_MFMlarge}b, respectively.
The measurement was started from an ME-poled state with ($+E_0$,$+H_0$) fields in the $\mathbf{E}\perp\mathbf{H}$; $\mathbf{E},\mathbf{H}\perp{c}$ configuration.
Magnitudes of the poling fields were $E_0$=3\,MV/m and $H_0$=4\,kOe.
Figure~\ref{ytypeS_MFMlarge}c shows the changes between the switched and initial images as a difference between the MFM phases.
The two MFM data were brought to complete overlapping using the same feature in their corresponding topography images as a point of reference (origin in Fig.~\ref{ytypeS01c}c).
Although Figs.~\ref{ytypeS_MFMlarge}a and \ref{ytypeS_MFMlarge}b look rather similar at a first glance, Fig.~\ref{ytypeS_MFMlarge}c highlights that there are many differences all over the measurement area, demonstrating that magnetic domains respond to the applied $E$ field, namely, there are many $P$-$M$ clamped DWs.

Figure~\ref{ytypeS_MFMsmall}a shows the changes in the magnetic domain pattern for four consecutive applications of the $E$ field in a particular area with 1$\times$2\,$\mu$m$^2$ dimensions (see Fig.~\ref{ytypeS01c}c).
As described in the main text, the application of $E$ field causes several changes in the magnetic domain pattern.
Firstly, the DW parallel to the $ab$ plane shifts along the $c$ axis.
The low anisotropy within the $ab$ plane suggests that the net $M$ is gradually twisted along the $c$ axis within the DW.
Secondly, a magnetization was reversed via the propagation of a DW within the $ab$ plane.
As the net $M$ of the FE3 phase is strongly confined to the $ab$ plane, the local $M$ is twisted in a cycloidal manner within this type of DW.
This type of DW is significantly wider ($\sim$500\,nm, see Fig.~\ref{ytype05}c in the main text) than the DWs between the stripe domains along the $c$ axis.
Moreover, one example for reversible domain switching can be found in region R4 from the 2nd to the 4th switches in Fig.~\ref{ytypeS_MFMsmall}a.
Both types of DW propagation contributes to the observed $M$ reversal, and the former can change the thickness along the $c$ axis of the domains, while the latter process can vary the area within the $ab$ plane.

While there are many changes in the magnetic domain pattern, reversed magnetization is not so large at room temperature in the present compound.
A possible explanation for this is provided by an example shown in Figs.~\ref{ytypeS_MFMsmall}b and \ref{ytypeS_MFMsmall}c.
In region R3, the MFM phase takes zero value in a broad area after the first application of $E$ field, and remains unchanged after the following switches.
Figure~\ref{ytypeS_MFMsmall}d schematically illustrates what happens in this region.
The DW between the FE3$+$ (positive magnetization) and FE3$-$ (negative magnetization) domains is turned into the ALC/PS state.
It is possible that the ALC/PS phase never returns to the FE3 phase in response to $E$ field, therefore the $M$-$E$ controllability is gradually deteriorated.


	\begin{center}
    \begin{figure*}[p]
 
    \includegraphics[width=7truecm]{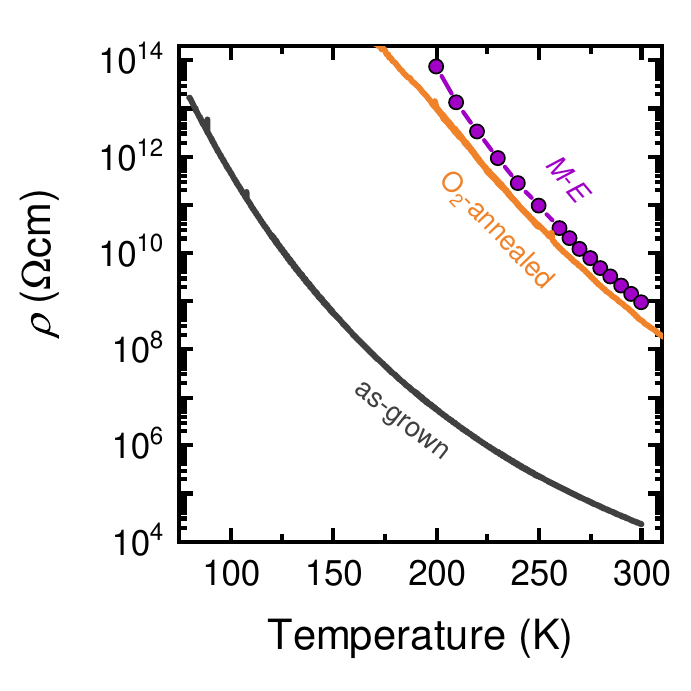}
    \caption{\textbf{ Electrical resistivity of BSCFAO before and after O$_2$ annealing.}
    Resistivity measured in temperature sweep under constant voltage, before and after the high-pressure O$_2$ annealing, are presented by black and red curves.
    The orange symbols represent the resistivity obtained from the $I$-$V$ curves at fixed temperatures during the measurement of the $M$-$E$ loops.}
    \label{ytypeS01a}
    \end{figure*}
    \end{center}

    \begin{center}
    \begin{figure*}[p]
 
    \includegraphics[width=16truecm]{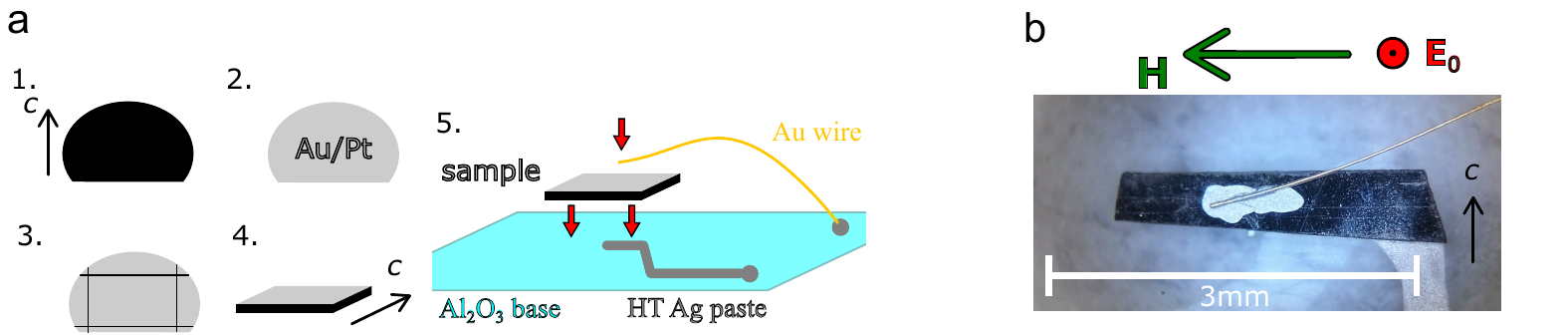}
    \caption{\textbf{ Sample preparation procedure for the $M$-$E$ measurements.}
    \textbf{a}, Schematic illustration of the sample preparation steps for $M$-$E$ measurements. Polished plate with $\sim$70\,$\mu$m thickness (1) was covered with Au/Pt electrodes (2), then the edges were cut off (3-4). The cut pieces were fixed to sapphire plate with silver paste and Au wire (5). 
    \textbf{b}, A photograph of the sample prepared for $M$-$E$ measurements. Magnetic field was applied along the lateral direction of the sample, while $E$ field was applied perpendicular to the plate.}
    \label{ytypeS01b}
    \end{figure*}
    \end{center}

    \begin{center}
    \begin{figure*}[p]
 
    \includegraphics[width=16truecm]{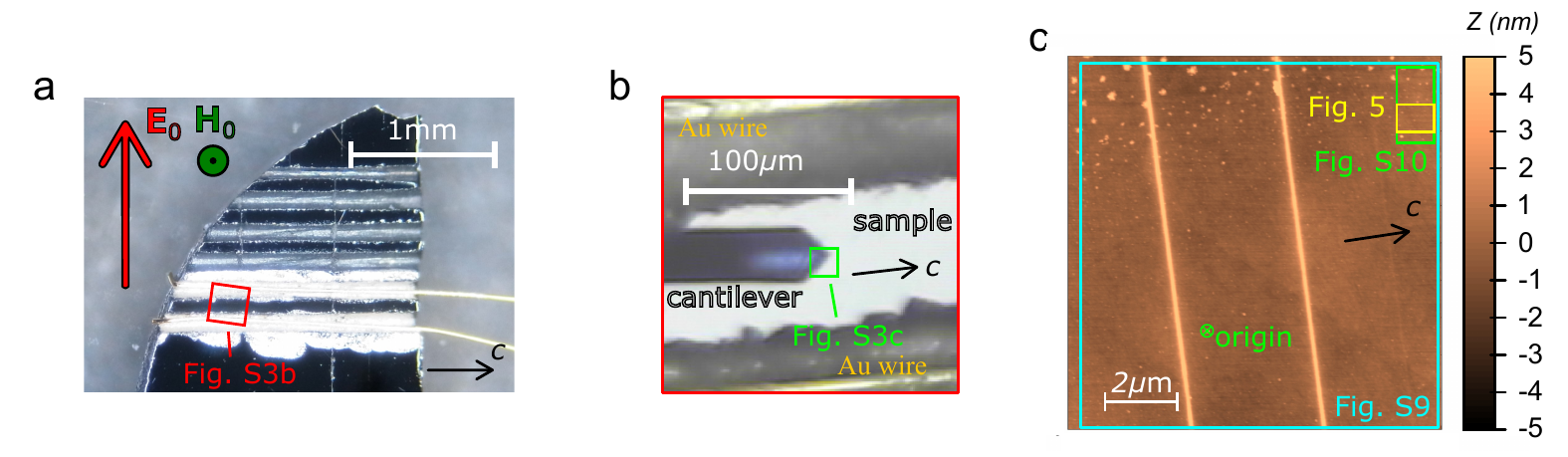}
    \caption{\textbf{ The sample for the MFM measurements.}
    \textbf{a}, Optical microscope image of the BSCFAO sample with an $ac$ surface prepared for the $E$-field dependent MFM experiments. $E$ and $H$ fields are applied within and perpendicular to the surface of the sample, respectively. Several grooves were cut so that they can be used to form electrodes.
    \textbf{b}, High magnification optical microscope image of the area of the measurement using the built-in camera of the MFM setup. The $E$-field dependent MFM images with 10$\times$10\,$\mu$m$^2$ dimension were taken on the area labeled by a green square.
    \textbf{c}, Topography image, which is measured simultaneously with the MFM images. This area was chosen due to the fixed features (two parallel cracks and particles), which are easily recognized and makes it reproducible to re-locate the frame of the MFM measurement to the same area. MFM data were analyzed using the same feature (a small particle) common in each topography images as an origin.}
    \label{ytypeS01c}
    \end{figure*}
    \end{center}

    \begin{center}
    \begin{figure*}[p]
 	
    \includegraphics[width=12truecm]{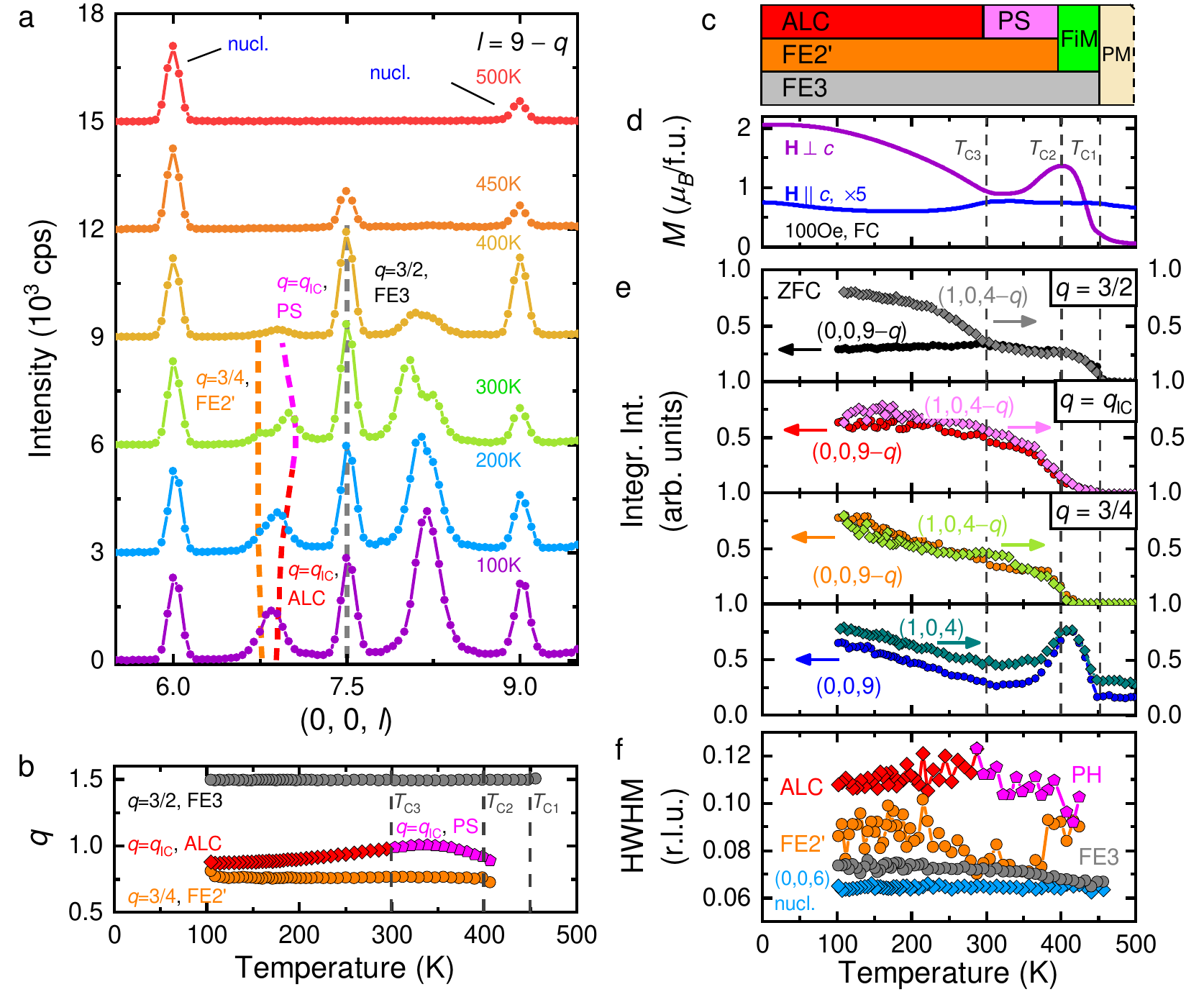}
    \caption{\textbf{ Determination of the magnetic phases based on the neutron diffraction profiles.}
    \textbf{a}, Temperature dependence of the neutron diffraction profiles measured along the $(0,0,l)$ line at $H$=0\,Oe.
    \textbf{b}, Temperature dependence of the wavenumbers of the FE3, FE2', ALC and PS phases.
    Panel \textbf{c} is the magnetic phase diagram for zero-field-cooling and panel \textbf{d} shows the temperature dependence of the magnetization (the same as in Fig.~\ref{ytype01} of the main text, re-plotted for better comparison). 
    \textbf{e}, The integrated intensities of the neutron diffraction peaks plotted against the temperature.
    The integrated intensities are taken along the $(0,0,l)$ and $(1,0,l)$ lines in the zero-field-cooled runs.
    In the paramagnetic phase ($T$=500\,K) only the two nuclear peaks at $l$=6 and $l$=9 can be observed.
    The FiM phase is signaled by the intensity change in the $(0,0,9)$ and $(1,0,4)$ reflections, while the FE3 phase, which is related to the magnetic peak with $q$=3/2 wavenumber ($l=9-q$), appears at $T$=450\,K.
    The magnetic reflections with commensurate $q$=3/4 and incommensurate $q_{\rm IC}$ wavenumbers, corresponding to the FE2' and ALC/PS phases, appear around $T$=400\,K.
    The FiM, FE2', and FE3 phases have ferrimagnetic component parallel to the net $M$, which contributes to the $(0,0,9)$ and $(1,0,4)$ peaks.
    The gradual increase below $T$=300\,K can be explained by an increase of volume fraction of the FE2' phase.
    \textbf{f}, Temperature dependence of the half widths at half maximums (HWHM) of the $(0,0,6)$ nuclear peak and the magnetic peaks with $q$=3/2, $q$=3/4, and $q$=$q_{\rm IC}$ wavenumbers, corresponding to the FE3, FE2' and ALC/PS phases, respectively.}
    \label{ytypeS02}
    \end{figure*}
    \end{center}

    \begin{center}
    \begin{figure*}[p]
 	
    \includegraphics[width=17truecm]{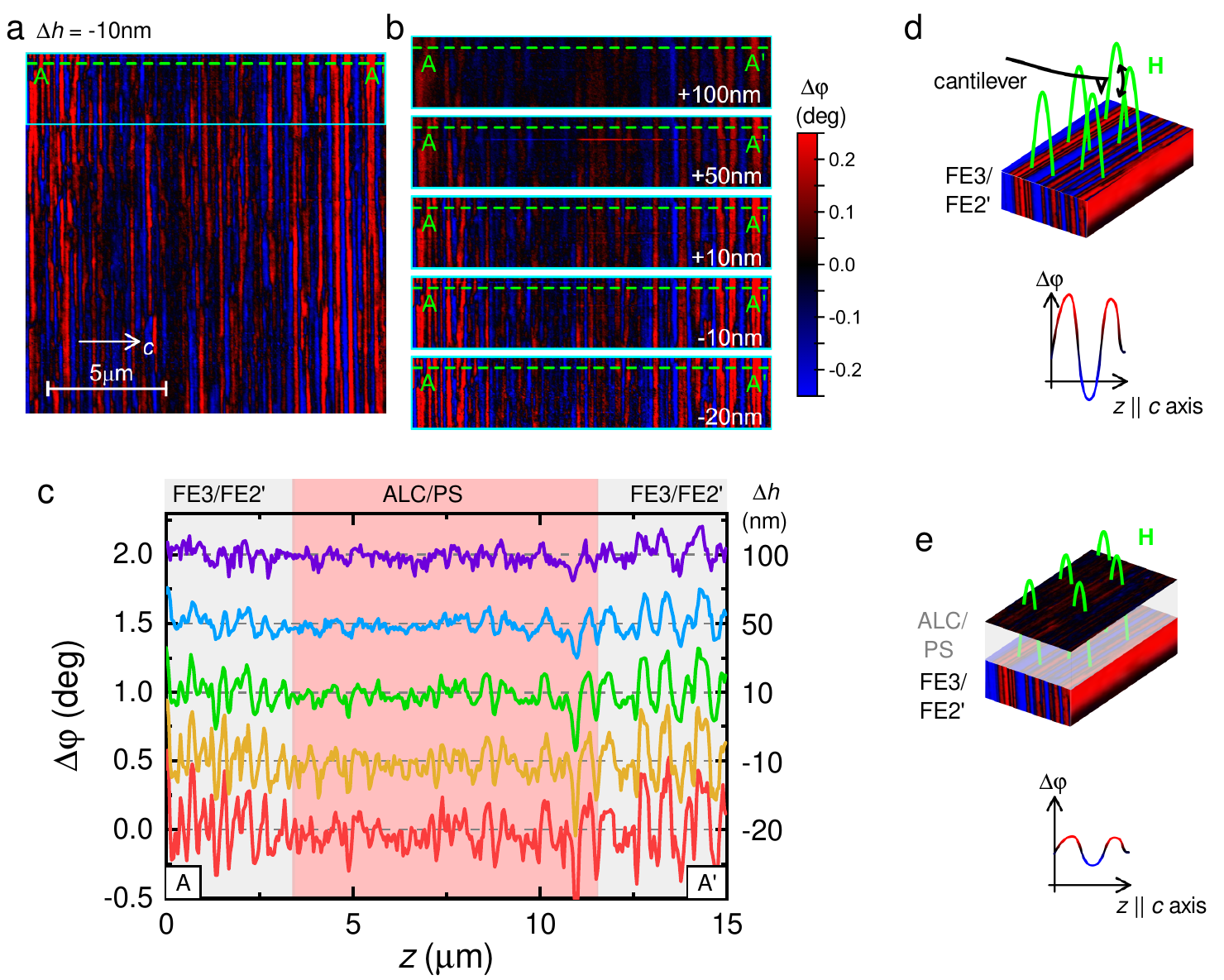}
    \caption{\textbf{ The magnetic domain structures as investigated by MFM measurements.}
    \textbf{a}, An MFM image of a region with 15$\times$15\,$\mu$m$^2$ dimensions showing the FE3/FE2' (left and right sides) and ALC/PS phases (middle) with large and small modulation of the MFM phase shifts, respectively.
    \textbf{b}, MFM measurements for the same area (surrounded by light-blue rectangle in panel a) with different cantilever distance from the surface. Negative $\Delta{h}$ means that the cantilever is closer to the sample surface during the MFM phase shift measurement than it was during the measurement of the topography.
    \textbf{c}, The MFM phase plotted along the line A-A' in panel a for various values of $\Delta{h}$.
    For each $\Delta{h}$, data are shifted by 0.5\,deg for the purpose of clarity.
    Schematic figure of two MFM measurements, when the cantilever is above an FE3/FE2' phase at the surface (\textbf{d}), or above an FE3/FE2' phase covered with an ALC/PS phase at surface (\textbf{e}).}
    \label{ytypeS_MFMhdep}
    \end{figure*}
    \end{center}

    \begin{center}
    \begin{figure*}[p]
 	
    \includegraphics[width=17truecm]{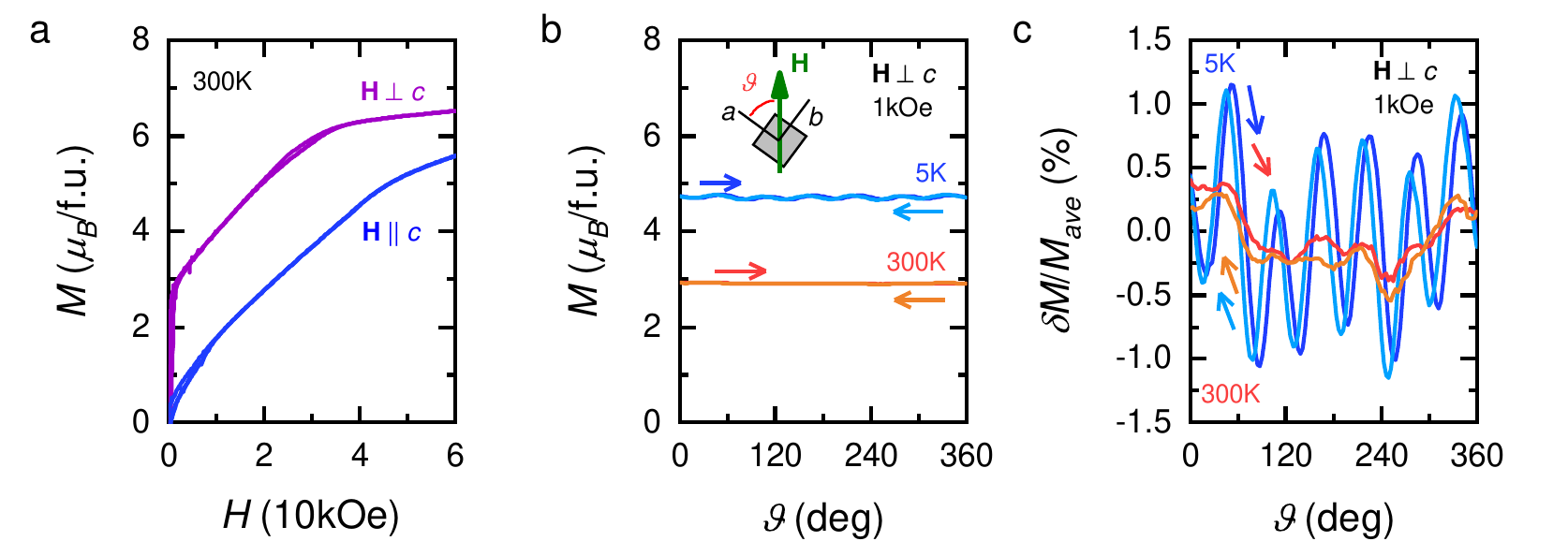}
    \caption{\textbf{ Magnetic anirotropy of BSCFAO.}
    \textbf{a}, $M$-$H$ curve of a BSCFAO sample at $T=$300\,K demonstrating the strong easy-plane anisotropy.
    \textbf{b}, Field-orientation dependence of the magnetic moment at $T=$300\,K and $T=$5\,K. Moderate $H$ field (1\,kOe) is rotated within the $ab$ plane, while the $M$ is measured.
    \textbf{c}, Magnetic anisotropy, relative to the average of the $M$, was calculated from the data shown in panel b.}
    \label{ytypeS_Manisotropy}
    \end{figure*}
    \end{center}

    \begin{center}
    \begin{figure*}[p]
 
    \includegraphics[width=16truecm]{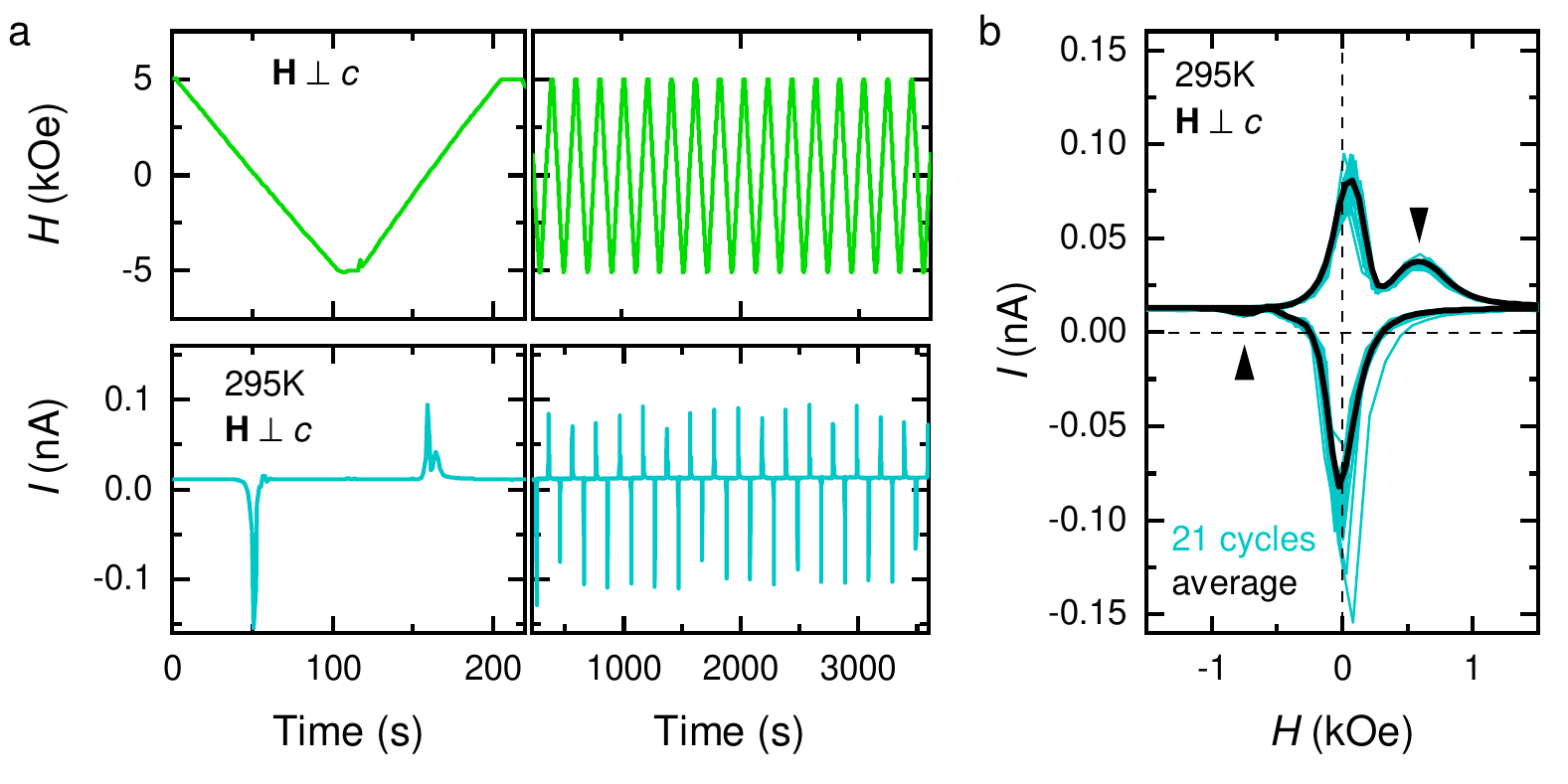}
    \caption{\textbf{ Measurement of the spin-induced polarization at $T$=295\,K.}
    \textbf{a}, Magnetic field dependence of the displacement current for the first cycle (left panel) and for subsequent cycles (right panel). Note the change in the scale of the horizontal axis.
    \textbf{b}, The measured displacement current  (orange) was averaged for 21 cycles to a single $I$-$H$ loop (blue). The averaged $I$ was integrated to obtain $P$, and the resultant $P$-$H$ curve is presented in Fig.~\ref{ytype03}.}
    \label{ytypeS03}
    \end{figure*}
    \end{center}

    \begin{center}
    \begin{figure*}[p]
 
    \includegraphics[width=8truecm]{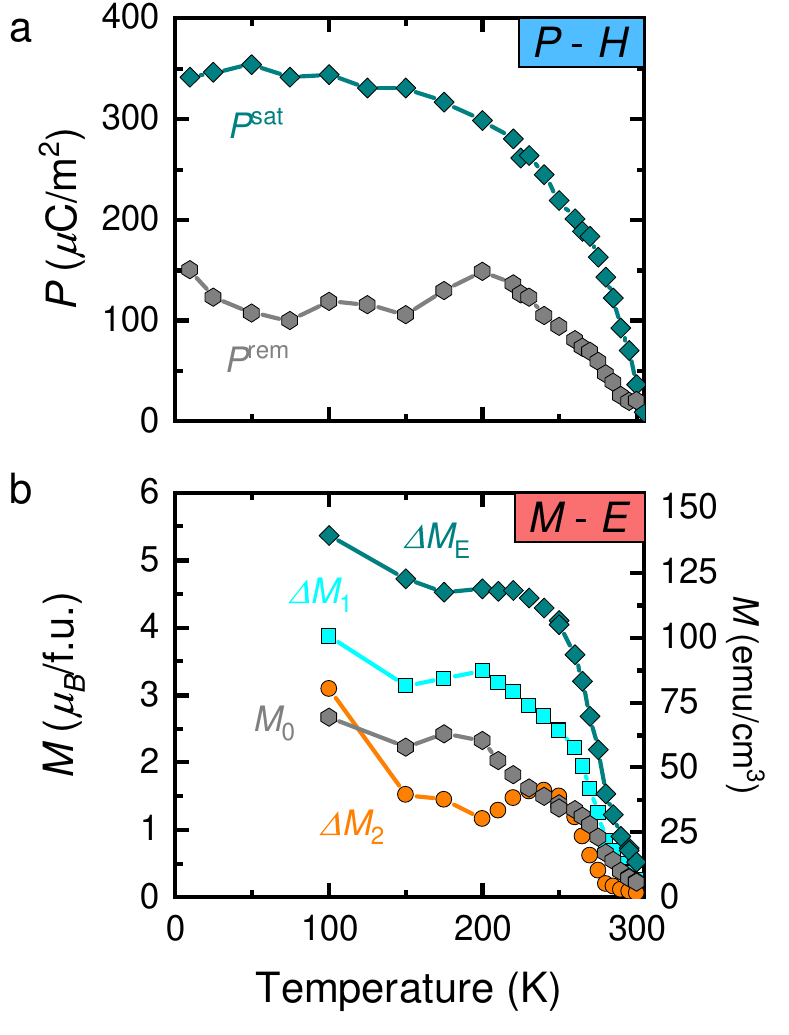}
    \caption{\textbf{ Characteristic quantities in quasi-static $P$-$H$ and $M$-$E$ measurements as a function of the temperature.}
    \textbf{a}, The saturation ($P^{\rm sat}$) and remanent ($P^{\rm rem}$) values obtained from $P$-$H$ measurement, have distinct temperature dependence from the $\Delta{P}_1$ and $\Delta{P}_2$ (see Fig.~\ref{ytype04}f) obtained from $P$-$E$ loops.
    \textbf{b}, The $M_0$, $\Delta{M}_1$ and $\Delta{M}_2$ values measured in quasi-static $E$ field sweep have similar temperature dependence to the corresponding values measured in pulsed $E$ field (see Fig.~\ref{ytype04}). $\Delta{M}_E$, the magnetization change between $\pm$\,5MV/m, follow similar temperature dependence to $P^{\rm sat}$. Due to its technological importance, $M$ values are presented also in the unit of emu/cm$^3$.}
    \label{ytypeS06}
    \end{figure*}
    \end{center}

    \begin{center}
    \begin{figure*}[p]
 
    \includegraphics[width=17truecm]{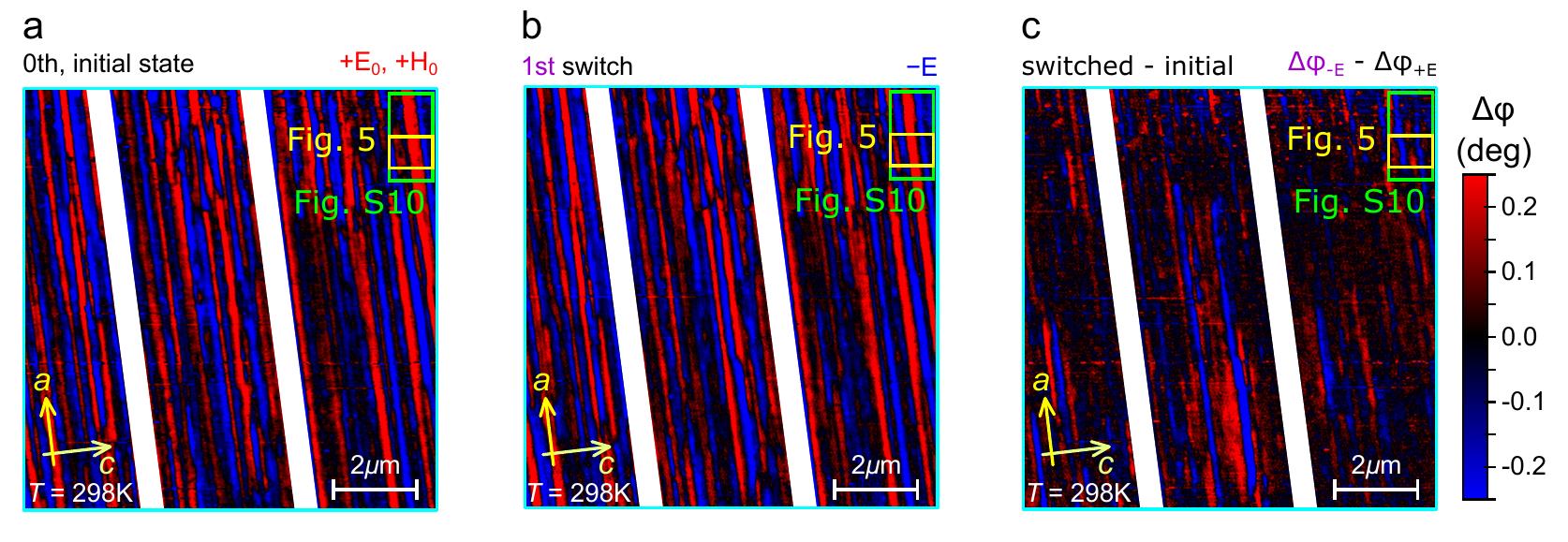}
    \caption{\textbf{ $E$ field manipulation of magnetic domains.}
    MFM images in a large area with 10$\times$10\,$\mu$m$^2$ dimensions shown in Fig.~\ref{ytypeS01c}c. 
    The measurement was started from an ME-poled state, with the poling fields applied in the $\mathbf{E}\perp\mathbf{H}$; $\mathbf{E},\mathbf{H}\perp{c}$ configuration.
    During the measurements both $E$ and $H$ fields were removed and the $E$ field switching was realized via the application of $-$3\,MV/m static field.
    Panels \textbf{a} and \textbf{b} show the MFM phase shifts before and after the first application of $E$ field of $-$3\,MV/m.
    The difference between the two images is shown in panel \textbf{c}.
    Slanted two blank areas in all the panels correspond to two scratches (see Fig.~q{ytypeS01c}c), and hence MFM data are not available.}
    \label{ytypeS_MFMlarge}
    \end{figure*}
    \end{center}

    \begin{center}
    \begin{figure*}[p]
 
    \includegraphics[width=17truecm]{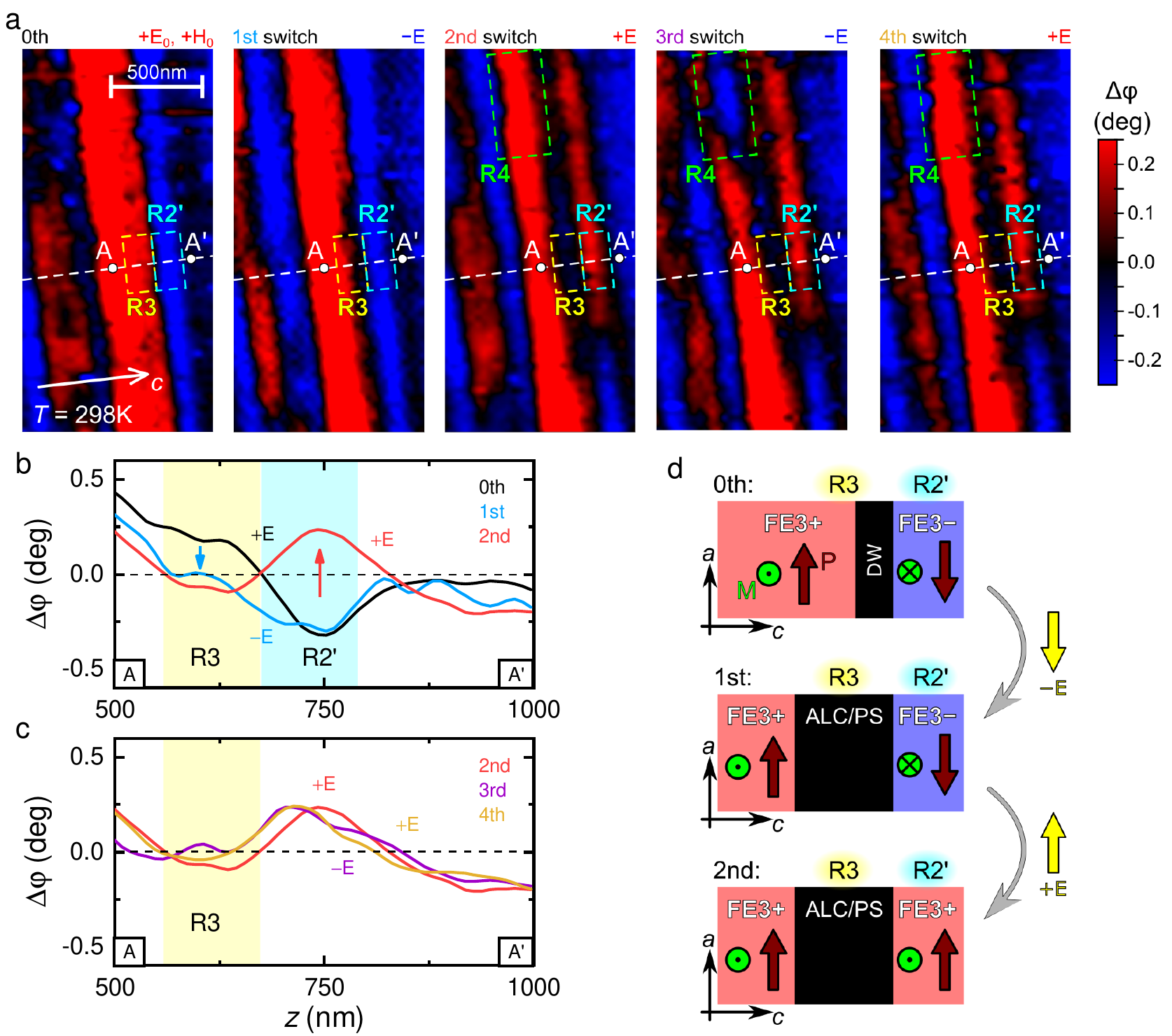}
    \caption{\textbf{ $E$ field manipulation of magnetic domains.}
    \textbf{a}, Evolution of the magnetic domain pattern for four consecutive applications of the $E$ field.
    \textbf{b}-\textbf{c}, MFM phase along the line AA' before and after four applications of the $E$ field.
    \textbf{d}, Schematic illustration of the MFM data at region R3.
    For the first applications of the $E$ field, the DW between FE3$+$ (positive magnetization) and FE3$-$ (negative magnetization) is turned into a weakly magnetic ALC/PS phase.
    For the second application of the $E$ field, the FE3$-$ domain is reversed to FE3+ at R2', which is caused by the propagation of DW along the $ab$-plane.}
    \label{ytypeS_MFMsmall}
    \end{figure*}
    \end{center}

\end{document}